\documentclass[journal]{IEEEtran}

\usepackage{graphicx, graphics, theorem, times, amsfonts, amsmath, amssymb, cite, caption}
\usepackage{tikz}
\usetikzlibrary{shapes,arrows}
\tikzstyle{phantom vertex} = [ ellipse, 
                               anchor = center, 
                               minimum height = 1*\unit, 
                               minimum width  = 1*\unit,
                               inner sep=0pt,
                               anchor=center]
\tikzstyle{red vertex}   = [black, fill = red!20,   phantom vertex, draw]
\tikzstyle{black vertex} = [black, fill = black!20, phantom vertex, draw]
\tikzstyle{blue vertex}  = [black, fill = blue!20,  phantom vertex, draw]
\tikzstyle{green vertex} = [black, fill = green!20,  phantom vertex, draw]
\tikzstyle{yellow vertex} = [black, fill = yellow!20,  phantom vertex, draw]
\tikzstyle{vertex}       = [draw, phantom vertex]

\tikzstyle{point} = [ellipse, inner sep=0pt, draw, fill=white, anchor = center,
                     minimum height = 0.05*\unit, minimum width  = 0.05*\unit]

\usepackage{pgfplots}
\usepackage{color}
\usepackage{needspace}

\usepackage{hyperref}
\usepackage{multirow}
\usepackage{rotating}
\usepackage{cleveref, subcaption}
\usepackage{booktabs}
\usepackage{bm}
\input{mysymbol.sty}

\newtheorem{lemma}{\hspace{0pt}\bf Lemma}
\newtheorem{proposition}{\hspace{0pt}\bf Proposition}

\newtheorem{theorem}{\hspace{0pt}\bf Theorem}
\newtheorem{corollary}{\hspace{0pt}\bf Corollary}

\newtheorem{remark}{\hspace{0pt}\bf Remark}

\def \d {\text{d}}
\def \sps {\text{sps}}
\def \diff {\text{diff}}
\newcommand{\QED}{\hfill\ensuremath{\blacksquare}}
\begin{document}
\title{Diffusion and Superposition Distances for \\ Signals Supported on Networks}
\author{Santiago Segarra, Weiyu Huang and Alejandro Ribeiro
\thanks{Work in this paper is supported by NSF CCF-1217963. The authors are with the Department of Electrical and Systems Engineering, University of Pennsylvania, 200 South 33rd Street, Philadelphia, PA 19104. Email: \{ssegarra, whuang, aribeiro\}@seas.upenn.edu.}}

\maketitle


%
\begin{abstract}
We introduce the diffusion and superposition distances as two metrics to compare signals supported in the nodes of a network. Both metrics consider the given vectors as initial temperature distributions and diffuse heat trough the edges of the graph. The similarity between the given vectors is determined by the similarity of the respective diffusion profiles. The superposition distance computes the instantaneous difference between the diffused signals and integrates the difference over time. The diffusion distance determines a distance between the integrals of the diffused signals. We prove that both distances define valid metrics and that they are stable to perturbations in the underlying network. We utilize numerical experiments to illustrate their utility in classifying signals in a synthetic network as well as in classifying ovarian cancer histologies using gene mutation profiles of different patients. We also reinterpret diffusion as a transformation of interrelated feature spaces and use it as preprocessing tool for learning. We use diffusion to increase the accuracy of handwritten digit classification.
\end{abstract}

%
\section{Introduction}\label{sec_introduction}

Networks, or graphs, are data structures that encode relationships between elements of a group and which, for this reason, play an important role in many disparate disciplines such as biology \cite{Bu03, Lieberman05} and sociology \cite{Newman06, Kleinberg99} where relationships between, say, genes, species or humans, are central. Often, networks have intrinsic value and are themselves the object of study. This is the case, e.g., when we are interested in distributed and decentralized algorithms in which agents iterate through actions that use information available either locally or at adjacent nodes to accomplish some sort of global outcome \cite{Kleinberg06, Kempe04, Lynch96}. Equally often, the network defines an underlying notion of proximity, but the object of interest is a signal defined on top of the graph. This is the matter addressed in the field of graph signal processing, where the notions of frequency and linear filtering are extended to signals supported on graphs \cite{Noble06, Miller10, Shuman13, Sandryhaila12, Narang11}. Examples of network-supported signals include gene expression patterns defined on top of gene networks \cite{Mittler04} and brain activity signals supported on top of brain connectivity networks \cite{Sporns11}. Indeed, one of the principal uses of networks of gene interactions is to determine how a change in the expression of a gene, or a group of genes, cascades through the network and alters the expression of other genes. Likewise, a brain connectivity network specifies relationships between areas of the brain, but it is the pattern of activation of these regions that determines the mental state of the subject.

In this paper we consider signals supported on graphs and address the challenge of defining a notion of distance between these signals that incorporates the structure of the underlying network. We want these distances to be such that two signals are deemed close if they are themselves close -- in the examples in the previous paragraph we have gene expression or brain activation patterns that are similar --, or if they have similar values in adjacent or nearby nodes -- the expressed genes or the active areas of the brain are not similar but they effect similar changes in the gene network or represent activation of closely connected areas of the brain. We define here the diffusion and superposition distances and argue that they inherit this functionality through their connection to diffusion processes.

Diffusion processes draw their inspiration from the diffusion of heat through continuous matter  \cite{Luikov68, Eckert87}. The linear differential equation that models heat diffusion can be extended to encompass dynamics through discrete structures such as graphs or networks \cite{Kondor02, Carrington05, Freidlin93, Szlam08, Smola03}. In the particular case of graphs, every node is interpreted as containing an amount of heat which flows from hot to cold nodes. The flow of heat is through the edges of the graph and such that the rate at which heat diffuses is proportional to a weight that defines the proximity between the nodes adjacent to the edge. Diffusion processes in graphs are often used in engineering and science because they reach isothermal configurations in steady state. Driving the network to an isothermal equilibrium is tantamount to achieving a consensus action \cite{Renetal05, Olfati07}, which, in turn, is useful in, e.g., problems in formation control \cite{Ren06} and flocking \cite{Tanneretal03}, as well as an important modeling tool in situations such as the propagation of opinions in social networks \cite{DeGroot74, Dittmer01, Segarra13}. 

In this paper we do not exploit the asymptotic, but rather the transient behavior of diffusion processes. We regard the given vectors as initial heat configurations that generate different diffused heat profiles over time. The diffusion and superposition distances between the given vectors are defined as the difference between these heat profiles integrated over time. The superposition distance compares the instantaneous difference between the two evolving heat maps and integrates this difference over time. The diffusion metric integrates each of the heat profiles over time and evaluates the norm of the difference between the two integrals. Both of these distances yield small values when the diffusion profiles are similar. This happens if the given vectors themselves are close or if they have similar values at nodes that are linked by edges with high similarity values. 

\subsection{Contributions and summary}

Besides the definition of the superposition and diffusion distances, the contributions of this paper are: (i) To prove that the superposition and diffusion distances are valid metrics in the space of vectors supported on a given graph. (ii) To show that both distances are well behaved with respect to small perturbations in the underlying network. (iii) To illustrate their ability to identify vectors that are similar only after the network structure is accounted for. (iv) To demonstrate their value in two practical scenarios; the classification of ovarian cancer types from gene mutation profiles and the classification of handwritten arabic digits.

We begin the paper with a brief introduction of basic concepts in graph theory and metric geometry followed by a formal description of diffusion dynamics in networks (Section \ref{sec_preliminaries}). This preliminary discussion provides the necessary elements for a formal definition of the superposition and diffusion distances. In Section \ref{sec_superposition_distance} we define the superposition distance between two signals with respect to a given graph and a given input norm. To determine this distance the signals are diffused in the graph, the input norm of their difference is computed for all times, and the result is discounted by an exponential factor and integrated over time. We show that the superposition distance is a valid metric between vectors supported in the node set of a graph. 

The diffusion distance with respect to a given graph and a given input norm is introduced in Section \ref{sec_diffusion_distance} as an alternative way of measuring the distance between two signals in a graph. In this case the diffused signals are also exponentially discounted and integrated over time but the input norm is taken after time integration. The diffusion distance is shown to also be a valid metric in the space of signals supported on a given graph and is further shown to provide a lower bound for the superposition distance. Different from the superposition distance, the diffusion distance can be reduced to a closed form expression with a computational cost that is dominated by a matrix inversion. The superposition distance requires numerical integration of the time integral of the norm of a matrix exponential. 

We further address stability with respect to uncertainty in the specification of the network (Section \ref{sec_stability}). Specifically, we prove that when the input norm is either the 1-norm, the 2-norm, or the infinity-norm a small perturbation in the underlying network transports linearly to a small perturbation in the values of the superposition and diffusion distances. In Section \ref{sec_applications} we demonstrate that the diffusion and superposition distances can be applied to classify signals in graphs with better accuracy than comparisons that utilize traditional vector distances. We illustrate the differences using synthetic data (Section \ref{sec_three_clusters}) and establish the practical advantages through the classification of ovarian cancer histologies from gene mutation profiles of different patients (Section \ref{sec_cancer_histology}). In Section \ref{sec_feature_space_transformation}, we reinterpret diffusion as a method for data preprocessing in learning for cases where interrelations exist across features in the feature space. We show the benefit of this data preprocessing through the classification of handwritten digits. We offer concluding remarks in Section \ref{sec_conclusion}.

%
\section{Preliminaries}\label{sec_preliminaries}

We consider networks that are weighted, undirected, and symmetric. Formally, we define a network as a graph $G=(V, E, W)$, where $V=\{1,\ldots,n\}$ is a finite set of $n$ nodes or vertices, $E\subseteq V\times V$ is a set of edges defined as ordered pairs $(i,j)$, and $W: E \to \reals_{++}$ is a set of strictly positive weights $w_{ij}>0$ associated with each edge $(i,j)$. Since the graph is undirected, we must have that the edge $(i,j)\in E$ if and only if $(j,i)\in E$. Since the graph is also symmetric, we must have $w_{ij}=w_{ji}$ for all $(i,j)\in E$. The edge $(i,j)$ represents the existence of a relationship between $i$ and $j$ and we say that $i$ and $j$ are adjacent or neighboring. The weight $w_{ij}=w_{ji}$ represents the strength of the relationship, or, equivalently, the proximity or similarity between $i$ and $j$. Larger edge weights are interpreted as higher similarity between the border nodes. The graphs considered here do not contain self loops, i.e., $(x, x) \not\in E$ for any $x \in V$. 

We consider the usual definitions of the adjacency, Laplacian, and degree matrices for the weighted graph $G=(V, E, W)$; see e.g. \cite[Chapter 1]{Chung97}. The adjacency matrix $A \in \reals_+^{n \times n}$ is such that $A_{ij}=w_{ij}$ whenever $i$ and $j$ are adjacent, i.e., whenever $(i,j)\in E$ and such that for $(i,j)\notin E$ we have $A_{ij}=0$. The degree matrix $D\in\reals_+^{n\times n}$ is a diagonal matrix such that the $i$-th diagonal element $D_{ii}=\sum_j w_{ij}$ contains the sum of all the weights out of node $i$. The Laplacian matrix is defined as the difference $L:=D-A\in\reals^{n\times n}$. Since $D$ is diagonal and the diagonal of $A$ is null -- because $G$ does not have self loops -- the components of the Laplacian matrix are explicitly given by
\begin{equation}\label{eqn_def_graph_laplacian}
   L_{ij}:= \begin{cases}
               -A_{ij}             \quad\qquad \text{if} \quad i \neq j, \\
               \sum_{k=1}^n A_{ik} \quad  \text{if} \quad i = j.
            \end{cases}
\end{equation}
Observe that the Laplacian is positive semidefinite because it is diagonally dominant with positive diagonal elements.

%
\subsection{Metrics and norms}

Our goal in this paper is to define a metric to compare vectors defined on top of a graph. For reference, recall that for a given space $X$, a metric $d: X \times X \to \reals_+$ is a function from pairs of elements in $X$ to the nonnegative reals satisfying the following three properties for every $x, y, z \in X$:

\begin{mylist}
\item[{\it Symmetry:}] $d(x, y)=d(y, x)$.
\item[{\it Identity:}] $d(x, y)=0$ if and only if $x=y$.
\item[{\it Triangle inequality:}] $d(x, y) \leq d(x, z) + d(z, y)$.
\end{mylist}

A closely related definition is that of a norm. In this case we need to have a given vector space $Y$ and consider elements $v\in Y$. A norm $\| \cdot \|$ is a function $\| \cdot \| : Y \to \reals_+$ from $Y$ to the nonnegative reals such that, for all vectors $v, w \in Y$ and scalar constant $\beta$, it satisfies:

\begin{mylist}
\item[{\it Positiveness:}] $\|v\| \geq 0$ with equality if and only if $v = \vec{0}$.
\item[{\it Positive homogeneity:}] $\|\beta \,w \| = |\beta| \,\| w \|$.
\item[{\it Subadditivity:}] $\|v + w\| \leq \|v\| + \|w\|$.
\end{mylist}

Norms are more stringent than metrics because they require the existence of a null element with null norm. However, whenever a norm is defined on a vector space $Y$ it induces a distance in the same space as we formally state next \cite[Chapter 1]{Burago01}.

%
\begin{lemma}\label{lem_norm_induces_metric}
Given any norm $\|\cdot\|$ on some vector space $Y$, the function $d:Y\times Y\to\reals_+$ defined as $d(r, s):=\|r-s\|$ for all pairs $r, s \in Y$ is a metric.
\end{lemma}

%

In some of our proofs we encounter norms induced in the vector space of matrices $\reals^{n\times n}$ by norms defined in the vector space $\reals^n$. For a given vector norm $\| \cdot \|:\reals^n\to\reals_+$ the induced matrix norm $\| \cdot \|:\reals^{n\times n}\to\reals_+$ is defined as
\begin{equation}\label{eqn_definition_induced_matrix_norm}
   \| A \| := \sup_{\|x\|=1} \|Ax\|.
\end{equation}
I.e., the norm of matrix $A$ is equal to the maximum vector norm achievable when multiplying $A$ by a vector with unit norm. Apart from satisfying the three requirements in the definition of norms, induced matrix norms are compatible and submultiplicative \cite[Section 2.3]{Golub89}. That they are submultiplicative means that for any given pair of matrices $A, B \in \reals^{n\times n}$ the norm of the product does not exceed the product of the norms,
\begin{equation}\label{eqn_definition_submultiplicativity}
   \|AB\| \leq \|A\|\,\|B\|.
\end{equation}
That they are compatible means that for any vector $x\in\reals^n$ and matrix $A\in\reals^{n\times n}$ it holds,
\begin{equation}\label{eqn_definition_compatibility}
   \|Ax\| \leq \|A\|\,\|x\|.
\end{equation}
I.e., the vector norm of the product $Ax$ does not exceed the product of the norms of the vector $x$ and the induced norm of the matrix $A$.

%
\subsection{Diffusion dynamics}\label{sec_diffusion_dynamics}

Consider an arbitrary graph $G = (V, E, W)$ with Laplacian matrix $L$ and a vector $r=[r_1,\ldots,r_n]^T \in \reals^n$ where the component $r_i$ of $r$ corresponds to the node $i$ of $G$. For a given constant $\alpha > 0$, define the time-varying vector $r(t) \in \reals^n$ as the solution of the linear differential equation
\begin{equation}\label{eqn_diffusion_dynamics}
   \frac{\d \,r(t)}{\d \,t} = - \alpha \,L \,r(t), \qquad r(0) = r.
\end{equation}
The differential equation in \eqref{eqn_diffusion_dynamics} represents heat diffusion on the graph $G$ because $- L$ can be shown to be the discrete approximation of the continuous Laplacian operator used to describe the diffusion of heat in physical space \cite{Kondor02}. The given vector $r=r(0)$ specifies the initial temperature distribution and $r(t)$ represents the temperature distribution at time $t$. The constant $\alpha$ is the thermal conductivity and controls the heat diffusion rate. Larger $\alpha$ results in faster changing $r(t)$. The solution of \eqref{eqn_diffusion_dynamics} is given by the matrix exponential,
\begin{equation}\label{eqn_diffusion_dynamics_solution}
   r(t) = e^{-\alpha \,L \,t}\,r,
\end{equation}
as can be verified by direct substitution of $r(t)=e^{-\alpha Lt}r$ in \eqref{eqn_diffusion_dynamics}. The expression in \eqref{eqn_diffusion_dynamics_solution} allows us to compute the temperature distribution at any point in time given the initial heat configuration $r$ and the structure of the underlying network through its Laplacian $L$. Notice that as time grows, $r(t)$ settles to an isothermal equilibrium -- all nodes have the same temperature -- if the graph is connected.

It is instructive to rewrite \eqref{eqn_diffusion_dynamics} componentwise. If we focus on the variation of the $i$-th component of $r(t)$ and use the definition of $L$ in \eqref{eqn_def_graph_laplacian} to replace $L_{ik}=-A_{ik}$ and $L_{ii}=\sum_{k=1}^n A_{ik}$, it follows that \eqref{eqn_diffusion_dynamics} implies 
\begin{equation}\label{eqn_diffusion_dynamics_one_component}
   \frac{\d \,r_i(t)}{\d \,t} = \sum_{k=1}^n \alpha \,A_{ik} \left(r_k(t) - r_i(t)\right).
\end{equation}
Further recalling that $A_{ik}=0$ if $i$ and $k$ are not adjacent and that $A_{ik}=w_{ik}$ otherwise, we see that the sum in \eqref{eqn_diffusion_dynamics_one_component} entails multiplying each of the differences $r_k(t) - r_i(t)$ between adjacent nodes by the corresponding proximities $w_{ik}$. Thus, \eqref{eqn_diffusion_dynamics_one_component} is describing the flow of heat through edges of the graph. The flow of heat on an edge grows proportionally with the temperature differential $r_k(t) - r_i(t)$, but also with the proximity $w_{ik}$. Nodes with large proximity tend to equalize their temperatures faster, other things being equal. In particular, two initial vectors $r(0) = r$ and $ s(0) = s$ result in similar temperature distributions across time if they are themselves similar -- all $r_i$ and $s_i$ components are close --, or if they have similar initial levels at nodes with large proximity -- each component $r_i$ may not be similar to $s_i$ itself but similar to the component $s_j$ of a neighboring node for which the edge weight $w_{ij}$ is large. This latter fact suggests that the diffused vectors $r(t)$ and $s(t)$ define a notion of proximity between $r$ and $s$ associated with the underlying graph structure. We exploit this observation to define distances between signals supported on graphs in the following two sections.

%
\section{Superposition distance}\label{sec_superposition_distance}

Given an arbitrary graph $G=(V,E,W)$ with Laplacian matrix $L$, an input vector norm $\| \cdot \|$, and two signals $r,s \in \reals^n$ defined in the node space $V$, we define the superposition distance $d^{L}_{\sps}(r,s)$ between $r$ and $s$ as
\begin{equation}\label{eqn_definition_superposition_distance}
d^{L}_{\sps}(r,s) := \int_0^{+\infty} e^{-t} \,\left\| e^{- \alpha \,L \,t} (r - s)\right\| \,\d t,
\end{equation}
where $\alpha > 0$ corresponds to the diffusion constant in \eqref{eqn_diffusion_dynamics}. As we mentioned in the discussion following \eqref{eqn_diffusion_dynamics_one_component}, the distance $d^{L}_{\sps}(r,s)$ defines a similarity between $r$ and $s$ that incorporates the underlying network structure. Indeed, notice that the term inside the input norm corresponds to the difference $r(t)-s(t)$ between the vectors that solve \eqref{eqn_diffusion_dynamics} for initial conditions $r$ and $s$ [cf. \eqref{eqn_diffusion_dynamics_solution}]. This means that we are looking at the difference between the temperatures $r(t)$ and $s(t)$ at time $t$, which we then multiply by the dampening factor $e^{-t}$ and integrate over all times. These temperatures are similar if $r$ and $s$ are similar, or, if $r$ and $s$ have similar values at similar nodes. The dampening factor gives more relative importance to the differences between $r(t)$ and $s(t)$ for early times. This is necessary because after prolonged diffusion times the network settles into an isothermal equilibrium and the structural differences between $r$ and $s$ are lost. 

Exploiting the same interpretation, we can define the superposition norm of a vector $v \in \reals^n$ for a given graph with Laplacian matrix $L$ and a a given input norm $\|\cdot\|$ as
\begin{equation}\label{eqn_definition_superposition_norm}
\|v\|^L_{\sps} := \int_0^{+\infty} e^{-t} \,\left\| e^{- \alpha \,L \,t} v \right\| \,\d t.
\end{equation}
Although we are referring to $d^{L}_{\sps}(r,s)$ as the superposition distance between $r$ and $s$ and $\|v\|^L_{\sps}$ as the superposition norm of $v$ we have not proven that they indeed are valid definitions of distance and norm functions. As it turns out, they are. We begin by showing that $\|\cdot\|^L_{\sps}$ is a valid norm as we claim in the following proposition.

%
\begin{proposition}\label{prop_superposition_is_norm}
The function $\|\cdot\|^L_{\sps}$ in \eqref{eqn_definition_superposition_norm} is a valid norm on $\reals^n$ for every Laplacian $L$ and every input norm $\|\cdot\|$.
\end{proposition}

%
\begin{myproofnoname}
As stated in Section \ref{sec_preliminaries}, we need to show positiveness, positive homogeneity and subadditivity of $\|\cdot\|^L_{\sps}$. To show positive homogeneity, utilize the positive homogeneity of the input norm and the linearity of integrals to see that for every vector $v \in \reals^n$ and scalar $\beta$, it holds
\begin{align}\label{eqn_proof_sup_norm_positive_homo}
   \|\beta v\|^L_{\sps} \
      & =\ \int_0^{+\infty}e^{-t}\,\|e^{-\alpha\,L\,t}\beta v\|\,\d t   \nonumber \\
      & =\ |\beta| \int_0^{+\infty}e^{-t}\,\|e^{-\alpha\,L\,t}v\|\,\d t \nonumber \\
      & =\ |\beta| \|v\|^L_{\sps}.
\end{align}
In order to show subadditivity, pick arbitrary vectors $v, w \in \reals^n$ and use the subadditivity of the input norm $\|\cdot\|$ and the linearity of integrals to see that
\begin{align}\label{eqn_proof_sup_norm_subadditivity}
   \|v + w \|^L_{\sps} \
      & =\    \int_0^{+\infty} e^{-t} \,\| e^{- \alpha \,L \,t} (v + w)\|\,\d t \nonumber\\ 
      & \leq\ \int_0^{+\infty} e^{-t} \left( \| e^{- \alpha \,L \,t} v \|       
                + \| e^{- \alpha \,L \,t} w \| \right)  \d t                   \nonumber\\
      & =\    \|v\|^L_{\sps} + \|w\|^L_{\sps}, 
\end{align}
To show positiveness, first observe that for every $v \in \reals^n$ we have that $\| v\|^L_{\sps} \geq 0$ since for every time $t$ the argument of the integral in the definition \eqref{eqn_definition_superposition_norm} is the product of two nonnegative terms, an exponential and a norm which itself satisfies the positiveness property. The fact that $\| \vec{0} \|^L_{\sps} = 0$ is an immediate consequence of the definition \eqref{eqn_definition_superposition_norm}. Hence, we are only left to show that $\|v\|^L_{\sps} \neq 0$ for $v \neq 0$. To show this, it suffices to prove that the argument of the integral in \eqref{eqn_definition_superposition_norm} is strictly positive for every time $t$ which is implied by the fact that the matrix $e^{- \alpha \,L \,t}$ is strictly positive definite for every $t$. To see why this is true, notice that $- \alpha \,L \,t$ is a real symmetric matrix, thus, it is diagonalizable and has real eigenvalues. Consequently, the eigenvalues of $e^{- \alpha \,L \,t}$ are the exponentials of the eigenvalues of $- \alpha \,L \,t$ which are strictly positive. \end{myproofnoname}

%
If the superposition norm is a valid norm as shown by Proposition \ref{prop_superposition_is_norm} it induces a valid metric as per the construction in Lemma \ref{lem_norm_induces_metric}. This induced metric is the superposition distance defined in \eqref{eqn_definition_superposition_distance} as we show in the following corollary.

%
\begin{corollary}\label{cor_superposition_is_metric}
The function $d^L_{\sps}$ in \eqref{eqn_definition_superposition_distance} is a valid metric on $\reals^n$ for every Laplacian $L$ and every input norm $\|\cdot\|$.
\end{corollary}

%
\begin{myproofnoname}
Since $d^L_{\sps}(r, s) = \|r-s\|^L_{\sps}$ for all vectors $r, s \in \reals^n$ and $\| \cdot \|^L_{\sps}$ is a well-defined norm [cf. Proposition \ref{prop_superposition_is_norm}], Lemma \ref{lem_norm_induces_metric} implies that $d^L_{\sps}$ is a metric on $\reals^n$.
\end{myproofnoname}

%
The distance $d^L_{\sps}$ incorporates the network structure to compare two signals $r$ and $s$ supported in a graph with Laplacian $L$. As a particular case the edge set $E$ of the underlying graph $G$ may be empty. In this case, the Laplacian $L=\bbzero$ is identically null and we obtain from \eqref{eqn_definition_superposition_distance} that $d^\mathbf{0}_{\sps}(r,s)=\|r-s\|$. This is consistent with the fact that when no edges are present, the network structure adds no information to aid in the comparison of $r$ and $s$ and the superposition distance reduces to the standard distance induced by the input norm.

The computational cost of evaluating the superposition distance is significant in general. To evaluate $d^{L}_{\sps}(r,s)$ we approximate the improper integral in \eqref{eqn_definition_superposition_distance} with a finite sum and evaluate the norm of the matrix exponential $\left\| e^{- \alpha \,L \,t} (r - s)\right\|$ at the points required by the appropriate discretization. An alternative notion of distance for graph-supported signals  that is computationally more tractable comes in the form of the diffusion distance that we introduce in the next section.

%
\section{Diffusion distance}\label{sec_diffusion_distance}

Given an arbitrary graph $G=(V,E,W)$ with Laplacian $L$, an input vector norm $\| \cdot \|$ and two signals $r,s \in \reals^n$ defined in the node space $V$, the diffusion distance $d^L_{\diff}(r,s)$ between $r$ and $s$ is given by
\begin{equation}\label{eqn_definition_diffusion_distance}
   d^{L}_{\diff}(r,s) 
      := \left\| \int_0^{+\infty} e^{-t} \,e^{- \alpha \,L \,t} (r - s) \,\d t \right\|,
\end{equation}
with $\alpha > 0$ corresponding to the diffusion constant in \eqref{eqn_diffusion_dynamics}. As in the case of the superposition distance in \eqref{eqn_definition_superposition_distance}, the diffusion distance incorporates the graph structure in determining the proximity between $r$ and $s$ through the solutions $r(t)$ and $s(t)$ of \eqref{eqn_diffusion_dynamics} for initial conditions $r$ and $s$ [cf. \eqref{eqn_diffusion_dynamics_solution}]. The difference is that in the diffusion distance the input norm of the difference between $r(t)$ and $s(t)$ is taken {\it after} discounting and integration, whereas in the superposition distance the input norm is applied {\it before} discounting and integration. An interpretation in terms of heat diffusion is that the diffusion distance compares the total (discounted) energy that passes trough each node. The superposition distance compares the energy difference at each point in time and integrates that difference over time. Both are reasonable choices. Whether the superposition or diffusion distance is preferable depends on the specific application.

A definite advantage of the diffusion distance is that the matrix integral in \eqref{eqn_definition_diffusion_distance} can be resolved to obtain a closed solution that is more amenable to computation. To do so, notice that the primitive of the matrix exponential $e^{-t}e^{-\alpha L t} = e^{-(I+\alpha L) t}$ is given by $-(I+\alpha L)^{-1}e^{-(I+\alpha L) t}$ to conclude that \eqref{eqn_definition_diffusion_distance} is equivalent to
\begin{equation}\label{eqn_definition_diffusion_distance_2}
   d^{L}_{\diff}(r,s) = \left\| (I + \alpha L)^{-1} (r-s) \right\|.
\end{equation}
As in the case of the superposition distance of Section \ref{sec_superposition_distance} a vector norm can be defined based on the same heat diffusion interpretation used to define the distance in \eqref{eqn_definition_diffusion_distance}. Therefore, consider a given a graph with Laplacian $L$ and a given input norm $\|\cdot\|$ and define the diffusion norm of the vector $v \in \reals^n$ as
\begin{equation}\label{eqn_definition_diffusion_norm}
\|v\|^L_{\diff} := \left\| \int_0^{+\infty}  e^{-t} \,e^{- \alpha \,L \,t} v \,\d t \right\|  =  \left\| (I + \alpha L)^{-1} v  \right\|,
\end{equation}
where the second equality follows from the same primitive expression used in \eqref{eqn_definition_diffusion_distance_2}. 

The superposition distance is a proper metric and the superposition norm is a proper norm. We show first that $\|\cdot\|^L_{\diff}$ is a valid norm as we formally state next.

\begin{proposition}\label{prop_diffusion_is_norm}
The function $\|\cdot\|^L_{\diff}$ in \eqref{eqn_definition_diffusion_norm} is a valid norm on $\reals^n$ for every Laplacian $L$ and every input norm $\|\cdot\|$.
\end{proposition}
\begin{myproofnoname}
To prove the validity of $\|\cdot\|^L_{\diff}$ we need to show positiveness, positive homogeneity and subadditivity; see Section \ref{sec_preliminaries}. Positive homogeneity follows directly from the positive homogeneity of the input norm, i.e. for any vector $v \in \reals^n$ and scalar $\beta$ we have that
\begin{align}\label{eqn_pos_hom_diffusion}
\| \beta v \|^L_{\diff} &=  \| (I+ \alpha L)^{-1} \beta v\| \nonumber \\
&=  | \beta | \| (I+ \alpha L)^{-1} v\|  =  |\beta| \| v \|^L_{\diff}.
\end{align}
In order to show subadditivity, pick arbitrary vectors $v, w \in \reals^n$ and use the subadditivity of the input norm $\|\cdot\|$ to see that
\begin{align}\label{eqn_proof_diff_norm_subadditivity}
   \|v + w \|^L_{\diff} &= \| (I + \alpha L)^{-1} (v + w) \| \nonumber \\
   & \leq \| (I + \alpha L)^{-1} v \| + \| (I + \alpha L)^{-1} w \| \nonumber \\
   & = \|v \|^L_{\diff} + \| w \|^L_{\diff}.
\end{align}
Given the positiveness property of the input norm $\| \cdot \|$, to show positiveness of the diffusion norm $\|\cdot\|^L_{\diff}$ it is enough to show that $(I + \alpha L)^{-1} v \neq \vec{0}$ for all vectors $v \in \reals^n$ different from the null vector. This is implied by the fact that $(I + \alpha L)^{-1}$ is a positive definite matrix. To see why $(I + \alpha L)^{-1}$ is positive definite, first notice that $L$ is positive semidefinite as stated in Section \ref{sec_preliminaries}. Consequently, $\alpha L$ is also positive semidefinite since $\alpha > 0$ and $I+\alpha L$ is positive definite since every eigenvalue of $I+\alpha L$ is a unit greater than the corresponding eigenvalues of $\alpha L$, thus, strictly greater than 0. Finally, since inversion preserves positive definiteness, the proof is completed.
\end{myproofnoname}

From Proposition \ref{prop_superposition_is_norm} and Lemma \ref{lem_norm_induces_metric} it follows directly that that the diffusion distance defined in \eqref{eqn_definition_diffusion_distance} is a valid metric as we prove next.

\begin{corollary}\label{cor_diffusion_is_metric}
The function $d^L_{\diff}$ in \eqref{eqn_definition_diffusion_distance} is a valid metric on $\reals^n$ for every Laplacian $L$ and every input norm $\|\cdot\|$.
\end{corollary}
\begin{myproofnoname}
Since $d^L_{\diff}(r, s) = \|r-s\|^L_{\diff}$ for all vectors $r, s \in \reals^n$ and $\| \cdot \|^L_{\diff}$ is a well-defined norm [cf. Proposition \ref{prop_diffusion_is_norm}], Lemma \ref{lem_norm_induces_metric} implies that $d^L_{\diff}$ is a metric on $\reals^n$.
\end{myproofnoname}

As in the case of the superposition norm and distance, the diffusion norm and distance reduce to the input norm and its induced distance when the set edge is empty. In that case we have  $L=\mathbf{0}$ and it follows from the definitions in \eqref{eqn_definition_diffusion_norm} and \eqref{eqn_definition_diffusion_distance} that $\|v\|^L_{\diff}= \|v\|^\bbzero_{\diff}  = \|v\|$ and that  $d^{L}_{\diff}(r, s) = d^{\mathbf{0}}_{\diff}(r, s)=\|r-s\|$. 

The superposition and diffusion distance differ in the order in which the input norm and time integral are applied. It is therefore reasonable to expect some relationship to hold between their values. In the following proposition we show that the diffusion distance is a lower bound for the value of the superposition distance.

\begin{proposition}\label{prop_superposition_greater_diffusion}
Given any graph $G=(V, E, W)$ with Laplacian $L$, any two signals $r, s \in \reals^n$ defined in $V$ and any input vector norm $\| \cdot \|$, the diffusion distance $d^L_{\diff}(r,s)$ defined in \eqref{eqn_definition_diffusion_distance} is a lower bound on the superposition distance $d^L_{\sps}(r, s)$ defined in \eqref{eqn_definition_superposition_distance}
\begin{equation}\label{eqn_inequality_superposition_diffusion}
d^L_{\sps}(r, s) \geq d^L_{\diff}(r,s).
\end{equation}
\end{proposition}
\begin{myproof}
Since the exponential $e^{-t}$ in \eqref{eqn_definition_superposition_distance} is nonnegative, we may replace it with its absolute value to obtain
\begin{align}\label{eqn_proof_superposition_greater_diffusion_010}
d_{\sps}(r, s) & = \int_0^{+\infty} |e^{-t}| \,\| e^{- \alpha \,L \,t} (r - s)\| \,\d t \nonumber \\
& = \int_0^{+\infty}  \| e^{-t} e^{- \alpha \,L \,t} (r - s)\| \,\d t,
\end{align}
where we used the positive homogeneity property of the input norm to write the second equality. Further using the subadditivity property of the input norm we may write
\begin{align}\label{eqn_proof_superposition_greater_diffusion_020}
d_{\sps}(r, s) \geq \left\| \int_0^{+\infty}  e^{-t} e^{- \alpha \,L \,t} (r - s) \,\d t \,\right\|.
\end{align}
The right hand side of \eqref{eqn_proof_superposition_greater_diffusion_020} is the definition of the diffusion distance $d_{\diff}(r, s)$ in  \eqref{eqn_definition_diffusion_distance}. Making this substitution in \eqref{eqn_proof_superposition_greater_diffusion_020} yields \eqref{eqn_inequality_superposition_diffusion}. \end{myproof}

For applications in which the superposition distance is more appropriate, the diffusion distance is still valuable because, as it follows from Proposition \ref{prop_superposition_greater_diffusion}, it can be used as a lower bound on the superposition distance. This lower bound is useful because computing the diffusion distance is less expensive than computing the superposition distance.

\begin{figure}
\centering
\centerline{\def \thisplotscale {0.56}
\def \unit {\thisplotscale cm}

{\small
\begin{tikzpicture}[-stealth, scale = \thisplotscale]

    

    \node [red vertex, minimum height=0.6cm, minimum width=0.6cm] at (0,0) (1) {$x_1$};
    \node [blue vertex, minimum height=0.6cm, minimum width=0.6cm] at (0,-2) (2) {$x_2$};    
    \node [blue vertex, minimum height=0.6cm, minimum width=0.6cm] at (0,-4) (3) {$x_3$};
    \node [blue vertex, minimum height=0.6cm, minimum width=0.6cm] at (2,-2) (4) {$x_4$};
    \node [blue vertex, minimum height=0.6cm, minimum width=0.6cm] at (4,-2) (5) {$x_5$};
    \node [green vertex, minimum height=0.6cm, minimum width=0.6cm] at (6,-2) (6) {$x_6$};
    \node [yellow vertex, minimum height=0.6cm, minimum width=0.6cm] at (8,-2) (7) {$x_7$};
    \node [blue vertex, minimum height=0.6cm, minimum width=0.6cm] at (10,-2) (8) {$x_8$};
    \node [blue vertex, minimum height=0.6cm, minimum width=0.6cm] at (12,-1) (9) {$x_9$};
    \node [blue vertex, minimum height=0.6cm, minimum width=0.6cm] at (12,-3) (10) {$x_{10}$};

    
   \path (1) edge [-,bend left=0, above, thick] node {$$} (2);
   \path (2) edge [-,bend left=0, above, thick] node {$$} (1);
   \path (2) edge [-,bend left=0, above, thick] node {$$} (3);
   \path (3) edge [-,bend left=0, above, thick] node {$$} (2);
   \path (1) edge [-,bend left=0, above, thick] node {$$} (4);
   \path (4) edge [-,bend left=0, above, thick] node {$$} (1);
   \path (2) edge [-,bend left=0, below, thick] node {$$} (4);
   \path (4) edge [-,bend left=0, below, thick] node {$$} (2);
   \path (3) edge [-,bend left=0, below, thick] node {$$} (4);
   \path (4) edge [-,bend left=0, below, thick] node {$$} (3);
   \path (4) edge [-,bend left=0, below, thick] node {$$} (5);
   \path (5) edge [-,bend left=0, below, thick] node {$$} (4);
   \path (5) edge [-,bend left=0, below, thick] node {$$} (6);
   \path (6) edge [-,bend left=0, below, thick] node {$$} (5);
   \path (6) edge [-,bend left=0, below, thick] node {$$} (7);
   \path (7) edge [-,bend left=0, below, thick] node {$$} (6);
   \path (8) edge [-,bend left=0, below, thick] node {$$} (7);
   \path (7) edge [-,bend left=0, below, thick] node {$$} (8);
   \path (8) edge [-,bend left=0, below, thick] node {$$} (9);
   \path (9) edge [-,bend left=0, below, thick] node {$$} (8);
   \path (8) edge [-,bend left=0, below, thick] node {$$} (10);
   \path (10) edge [-,bend left=0, below, thick] node {$$} (8);

\end{tikzpicture}
} }
\caption{Example of an underlying graph used to compute the superposition and diffusion distances. Three signals $r$, $g$ and $y$ are compared taking a value of 1 in the red, green, and yellow nodes respectively, and zero everywhere else.}
\label{fig_network_example_diffusion}
\end{figure}
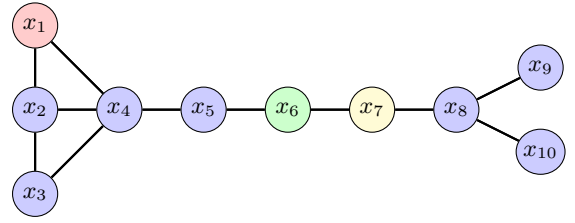
%

\subsection{Discussion}

\begin{figure*}
\centering

\begin{subfigure}{.33\textwidth}
  \centering
  \includegraphics[width=\textwidth]{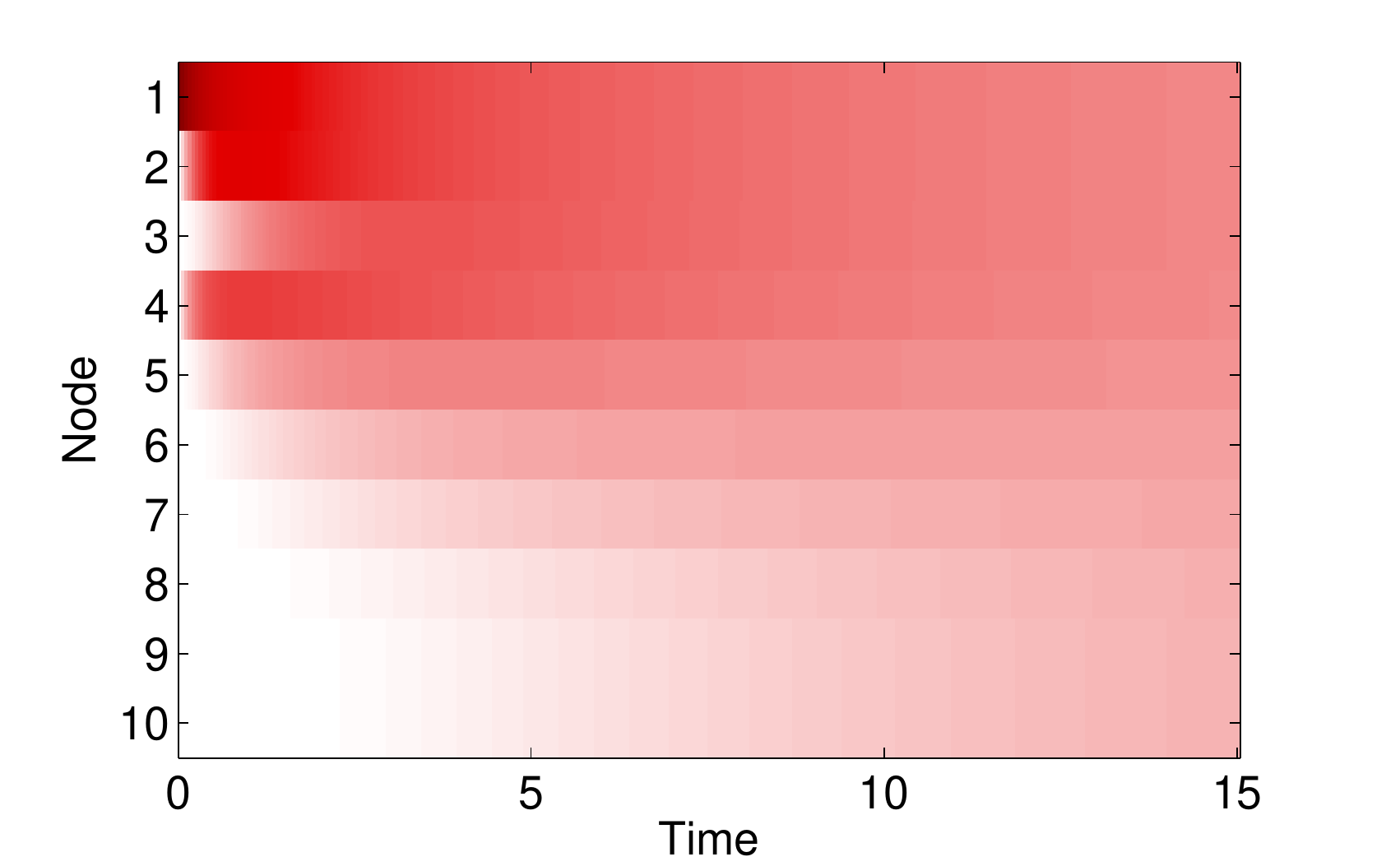}
  \caption{Diffusion of $r$}
  \label{fig_red_node_heat}
\end{subfigure}%
\begin{subfigure}{.33\textwidth}
  \centering
  \includegraphics[width=\textwidth]{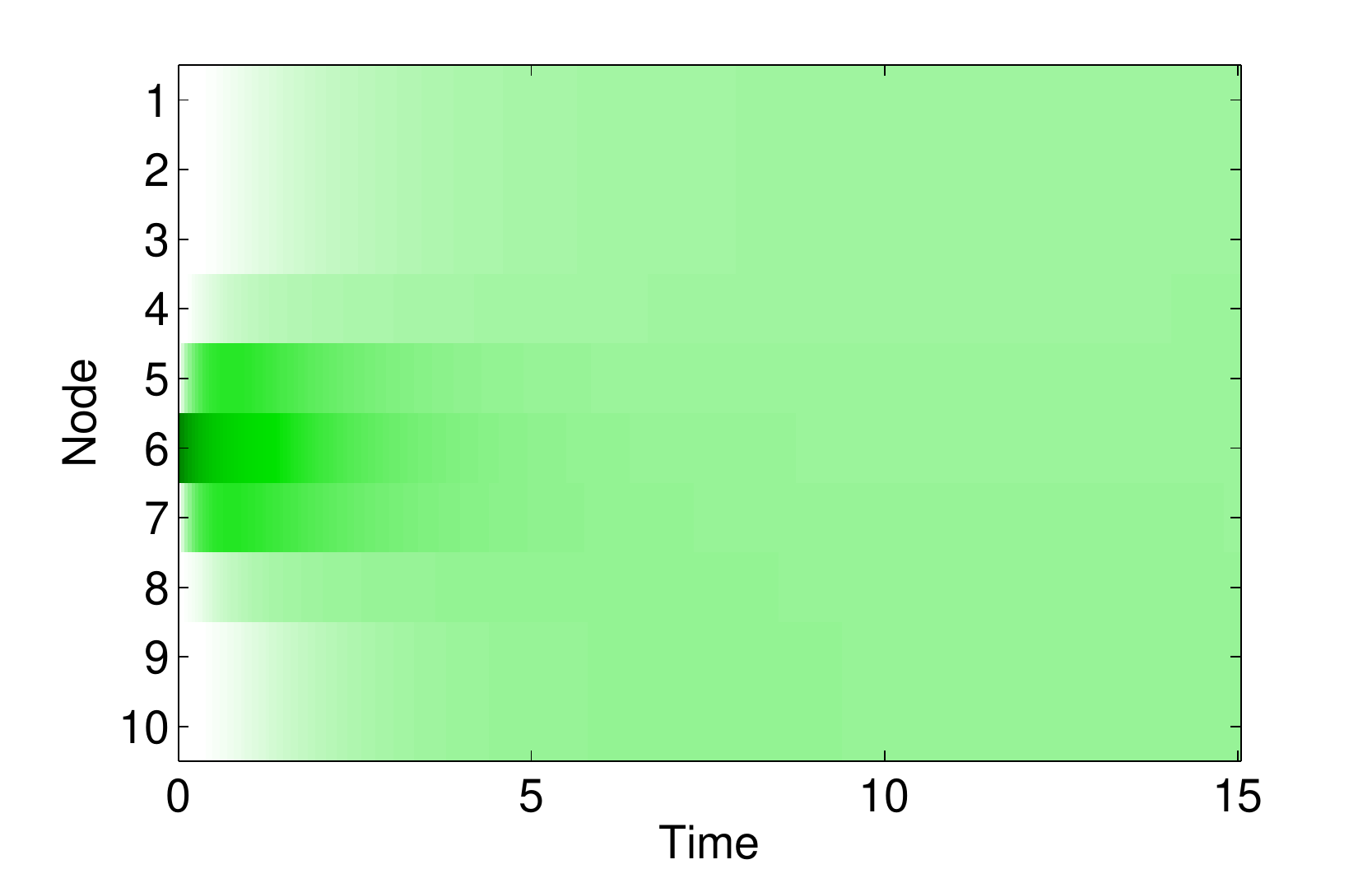}
  \caption{Diffusion of $g$}
  \label{fig_green_node_heat}
\end{subfigure}%
\begin{subfigure}{.33\textwidth}
  \centering
  \includegraphics[width=\textwidth]{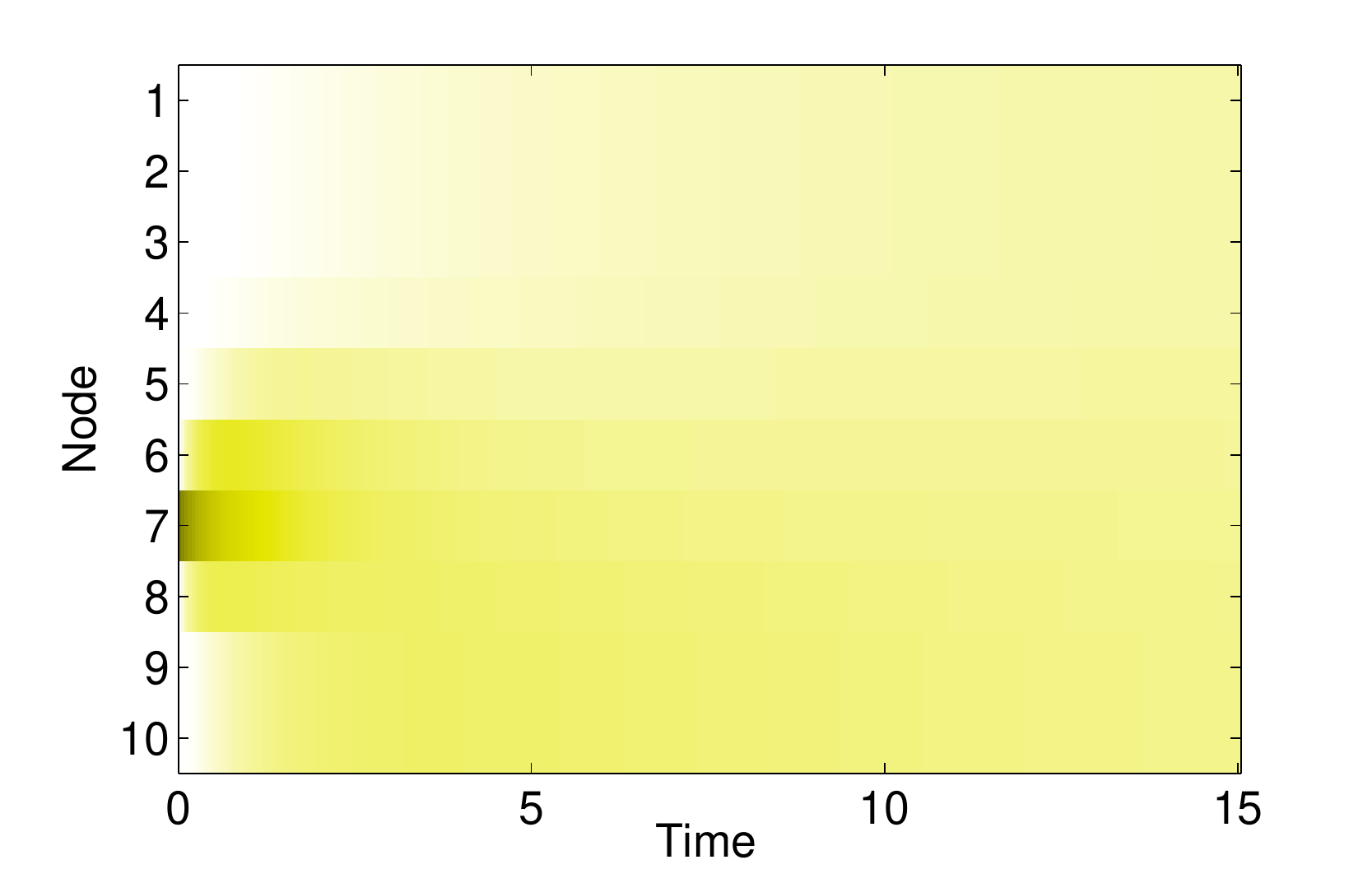}
  \caption{Diffusion of $y$}
  \label{fig_yellow_node_fig}
\end{subfigure}

\caption{Heat maps of the diffused signals for $r$, $g$, and $y$ as diffusion evolves for every node in the network in Figure \ref{fig_network_example_diffusion}. Darker colors represent stronger signals. The heat maps of $g$ and $y$ are more similar, entailing smaller diffusion and superposition distances.}
\label{fig_comparison_heat}
\end{figure*}

In order to illustrate the superposition and diffusion distances and their difference with the standard vector distances, consider the undirected graph in Figure \ref{fig_network_example_diffusion} where the weight of each undirected edge is equal to 1. Define three different vectors supported in the node space and having exactly one component equal to 1 and the rest equal to 0. The vector $r$ has its positive component for node $x_1$, colored in red, the vector $g$ has its positive for node $x_6$, colored in green, and the vector $y$ has its positive component for node $x_7$, colored in yellow.

For the traditional vector metrics, the distances between each of the vectors $r$, $g$ and $y$ is the same. In the case when, e.g., the $\ell_2$ distance is used as input metric, we have that $\| r - g \|_2 = \| g - y \|_2 = \| y - r\|_2 = \sqrt{2}$. In the case of the $\ell_1$ and $\ell_\infty$ distances we have that $\| r - g \|_1 = \| g - y \|_1 = \| y - r\|_1 = 2$ and $\| r - g \|_\infty = \| g - y \|_\infty = \| y - r\|_\infty = 1$. However, by observing the network in Figure \ref{fig_network_example_diffusion}, it is intuitive that signals $g$ and $y$ should be more alike than they are to $r$ since they affect nodes that are closely related. E.g., if we think of the vectors $r$, $g$ and $y$ as signaling faulty nodes in a communication network, it is evident that the impact of nodes $x_6$ and $x_7$ failing would disrupt the communication between the right and left components of the graph, whereas the failure of $x_1$ would entail a different effect. This intuition is captured by the diffusion and superposition distances. Indeed, if we fix $\alpha = 1$ and we use the $\ell_2$ norm as input norm to the diffusion distance, we have that the distance between the vectors that signal faults at $x_6$ and $x_7$ are [cf. \eqref{eqn_definition_diffusion_distance_2}]
\begin{equation}\label{eqn_example_diffusion_distance}
  d^L_{\diff}(g,y) = \| (I + L)^{-1} (g - y) \|_2 = 0.418,
\end{equation}
where $L$ is the Laplacian of the graph in Figure \ref{fig_network_example_diffusion}. However, the diffusion distances from these green and yellow vectors to the red vector that signals a fault at node $x_1$ are
\begin{align}\label{eqn_example_diffusion_distance_b}
   d^L_{\diff}(r,g) = \| (I + L)^{-1} (r - g) \|_2 = 0.664, \nonumber \\ 
   d^L_{\diff}(r,y) = \| (I + L)^{-1} (r - y) \|_2 = 0.698.
\end{align}
The distances in \eqref{eqn_example_diffusion_distance_b} are larger than the distance in \eqref{eqn_example_diffusion_distance} signaling the relative similarity of the $g$ and $y$ vectors with respect to the $r$ vector. The differences are substantial -- almost $60 \%$ increase --, thus allowing identification of $g$ and $y$ as somehow separate from $r$. Further observe that the distance between $r$ and $g$ is slightly smaller than the distance between $r$ and $y$. This is as it should be, because node $x_1$ is closer to node $x_6$ than to node $x_7$ in the underlying graph.

Repeating the exercise, but using the superposition distance instead [cf. \eqref{eqn_definition_superposition_distance}], we obtain that $d^L_{\sps}(r, g)=  0.701$, $d^L_{\sps}(r, y)=  0.742$, and $d^L_{\sps}(g, y)=  0.456$. Although the numbers are slightly different, the qualitative conclusions are the same as those obtained for the diffusion distance. We can tell that $g$ and $y$ are more like each other than they are to $r$, and we can tell that $g$ is slightly closer to $r$ than $y$ is. Also note that the diffusion distances are smaller than the superposition distances between the corresponding pairs, i.e., $d^L_{\sps}(r, g) \geq d^L_{\diff}(r, g)$, $d^L_{\sps}(r, y) \geq d^L_{\diff}(r, y)$, and $d^L_{\sps}(g, y) \geq d^L_{\diff}(g, y)$. This is consistent with the result in Proposition \ref{prop_superposition_greater_diffusion}.

To further illustrate the intuitive idea behind the diffusion and superposition distances, Figure \ref{fig_comparison_heat} plots the evolution of the diffused signals $r(t)$, $g(t)$ and $y(t)$ for each of the respective initial conditions $r$, $g$, and $y$. At time $t=0$ each of the signals is concentrated at one specific node. The signals are, as a consequence, equally different to each other. At very long times, the signals are completely diffused and therefore indistinguishable. For intermediate times, the signal distributions across nodes for the green and yellow signals are more similar than between the green and red or yellow and red signals. This difference between the evolution of the diffused signals results in different values for the superposition and diffusion distances.

\begin{remark}\label{rem_computational_diffusion}\normalfont
Computation of the diffusion distance using the closed form expression in \eqref{eqn_definition_diffusion_distance_2} requires the inversion of the $n \times n$ identity plus Laplacian matrix followed by multiplication with the difference vector $r-s$. The cost of this computation is of order $n^3$, but is much smaller when the matrix $L$ is sparse, as is typically the case. Further observe that most computations can be reused when computing multiple distances, because the vectors change, but the matrix inverse $(I+\alpha L)^{-1}$ stays unchanged.
\end{remark}


\section{Stability}\label{sec_stability}

\begin{figure}
\centering

\begin{subfigure}{.25\textwidth}
  \centering
  \includegraphics[width=\textwidth]{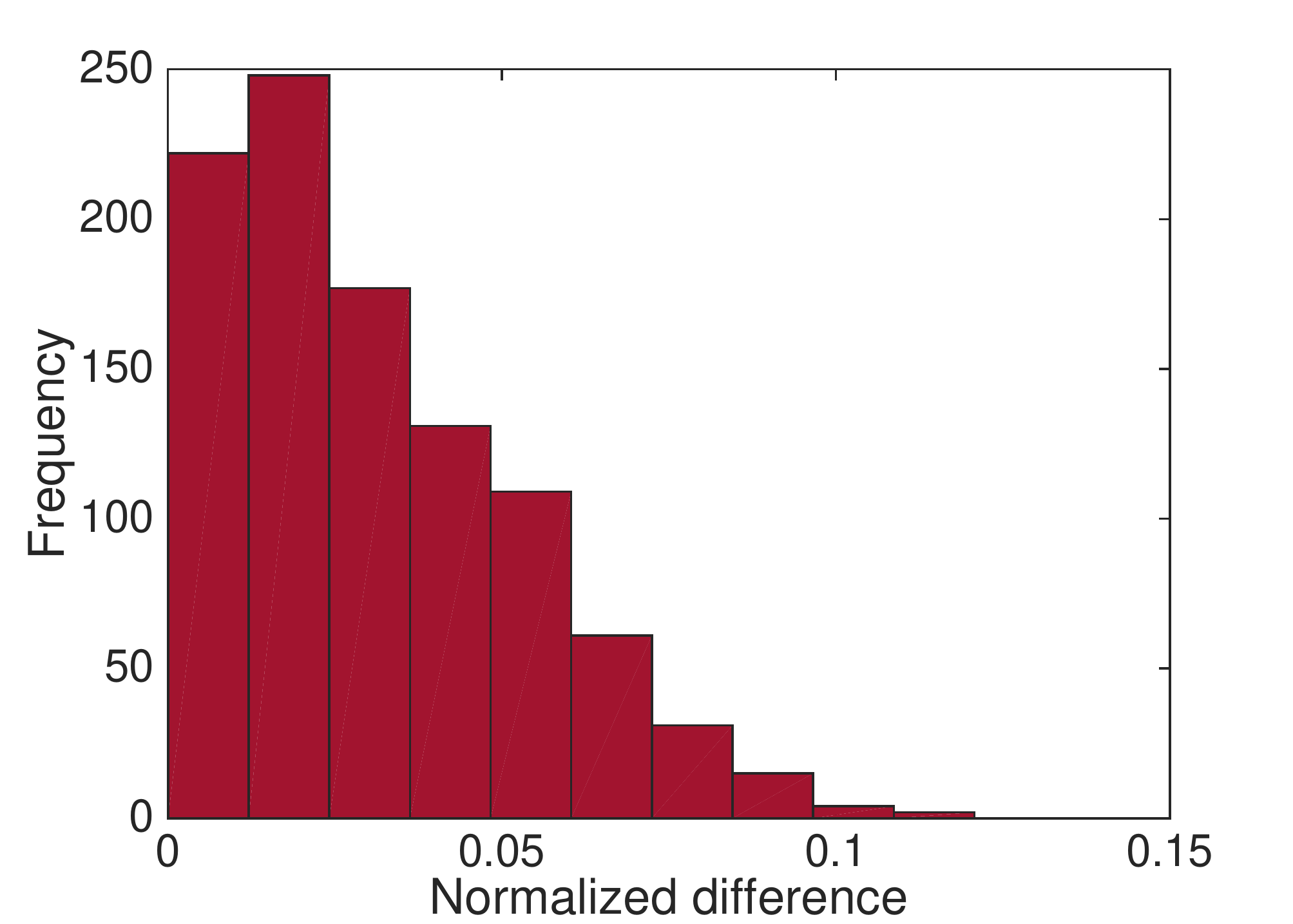}
  \caption{Diffusion distance}
  \label{fig_diff_stability_hist}
\end{subfigure}%
\begin{subfigure}{.25\textwidth}
  \centering
  \includegraphics[width=\textwidth]{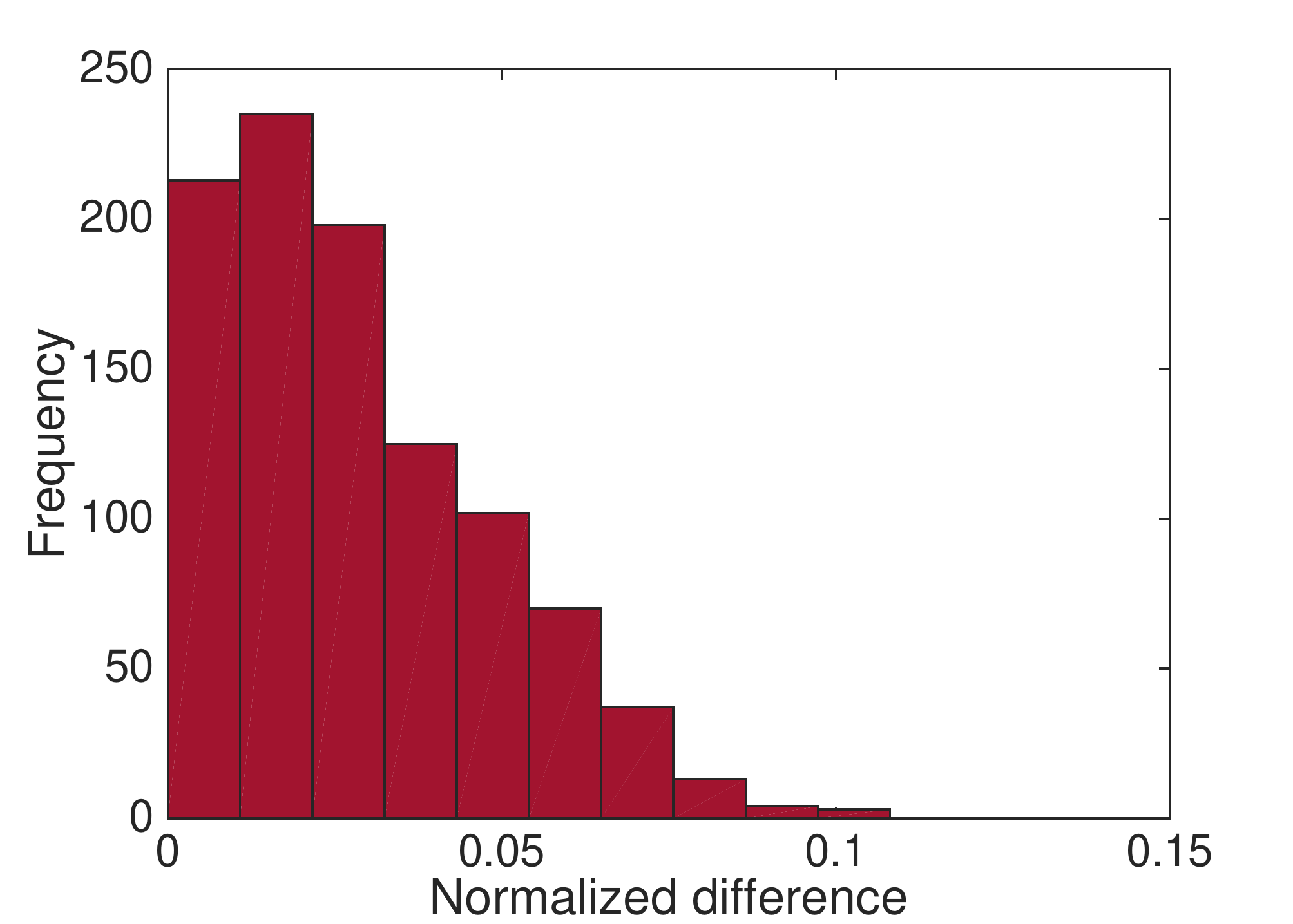}
  \caption{Superposition distance}
  \label{fig_sps_stability_hist}
\end{subfigure}

\caption{Histogram of the absolute value of the normalized difference, i.e. $|d^{L'}(g, r) -d^{L}(g, r)|/\|E\|_2$, for the diffusion and superposition distances. For this particular network and perturbations, the difference is considerably lower than the theoretical upper bound of 2.}
\label{fig_stability_hist}
\end{figure}

The superposition and diffusion distances depend on the underlying graphs through their Laplacian $L$. It is therefore important to analyze how a perturbation of the underlying network impacts both distances. We prove in this section that these distances are well behaved with respect to perturbations of the underlying graph. I.e., we show that if the network perturbation is small, the change in the diffusion and superposition distances is also small. We quantify the network perturbation as the matrix $p$-norm of the difference between the Laplacians of the original and perturbed networks. We focus our analysis on the most frequently used norms where $p \in \{1, 2, \infty\}$. We begin with a formal statement for the case of the superposition distance defined by  \eqref{eqn_definition_superposition_distance}.

%
\begin{theorem}\label{thm_spsnorm_stability}
Given any graph with Laplacian $L$, an input $\ell_p$ norm $\| \cdot\|_p$ with $p \in \{1, 2, \infty\}$, and bounded signals $s$ and $r$ on the network with $\|s\|_p \le \gamma$ and $\|r\|_p \le \gamma$, if we perturb the network such that the resulting Laplacian $L' = L + E$ where the perturbation $E$ is such that $\|E\|_p \le \epsilon \|L\|_p < 1$, then
\begin{align}\label{eqn_thm_spsnorm_stability}
      \left| d_{\sps}^{L'}(s,r) - d_{\sps}^{L}(s,r) \right| 
            \le 2 \gamma \|L\|_p \epsilon.
\end{align}
\end{theorem}
\begin{myproofnoname}
See Appendix \ref{app_stability_sps}.
\end{myproofnoname}

Theorem \ref{thm_spsnorm_stability} guarantees that for any two vectors, the difference between their superposition distances computed based on different underlying graphs is bounded by a term which is bilinear in a bound on the magnitude of the input vectors $\gamma$ and a bound on the difference between the Laplacians of both underlying graphs $ \|E\|_p \leq \epsilon \|L\|_p$. This implies that vanishing perturbations on the underlying network have vanishing effects on the distance between two signals defined on the network. 

Similarly to the case of the superposition distance, perturbations have limited effect on the diffusion metric defined in \eqref{eqn_definition_diffusion_distance} as shown next.

%
\begin{theorem}\label{thm_diffnorm_stability}
For the same setting described in Theorem \ref{thm_spsnorm_stability}, we have that
\begin{align}\label{eqn_thm_diffnorm_stability}
      \left|d_{\diff}^{L'}(s,r) - d_{\diff}^{L}(s,r) \right| \le 2 \gamma \|L\|_p\epsilon + o(\epsilon).
\end{align}
\end{theorem}
\begin{myproofnoname}
See Appendix \ref{app_stability_diff}.
\end{myproofnoname}

\begin{figure*}
\centering

\begin{subfigure}{.35\textwidth}
  \centering

\def \thisplotscale {0.45}
\def \unit {\thisplotscale cm}

\tikzstyle{node1} = [red vertex,
                    minimum height = 0.35*\unit, 
                    minimum width  = 0.35*\unit]
\tikzstyle{node2} = [green vertex,
                    minimum height = 0.35*\unit, 
                    minimum width  = 0.35*\unit]
\tikzstyle{node3} = [blue vertex,
                    minimum height = 0.35*\unit, 
                    minimum width  = 0.35*\unit]

\begin{tikzpicture}[x = 1*\unit, y=1*\unit, font=\tiny]

   
   \coordinate (center) at (0,0);
   \def \radius  {2}
   \def \nrnodes {9}
   
   \path [draw] (center) ++ ( 4 *360/\nrnodes:\radius) node [node1] (N1_1) {};
   \path [draw] (center) ++ ( 5 *360/\nrnodes:\radius) node [node1] (N1_2) {};
   \path [draw] (center) ++ ( 6 *360/\nrnodes:\radius) node [node1] (N1_3) {};
   \path [draw] (center) ++ ( 7 *360/\nrnodes:\radius) node [node1] (N1_4) {};
   \path [draw] (center) ++ ( 8 *360/\nrnodes:\radius) node [node1] (N1_5) {};
   \path [draw] (center) ++ ( 9 *360/\nrnodes:\radius) node [node1] (N1_6) {};
   \path [draw] (center) ++ ( 1 *360/\nrnodes:\radius) node [node1] (N1_7) {};
   \path [draw] (center) ++ ( 2 *360/\nrnodes:\radius) node [node1] (N1_8) {};
   \path [draw] (center) ++ ( 3 *360/\nrnodes:\radius) node [node1] (N1_9) {};
   
   \path [red, draw, line width = 0.5 pt] (center) (N1_1) arc (4 *360/\nrnodes: 5 *360/\nrnodes:\radius) ;
   \path (N1_1) edge [red, line width=1 pt] (N1_3);
   \path (N1_1) edge [red, line width=1 pt] (N1_5);
   \path (N1_1) edge [red, line width=0.5 pt] (N1_7);
   \path [red, draw, line width = 1.5 pt] (center) (N1_9) arc (12 *360/\nrnodes: 13 *360/\nrnodes:\radius) ;
   
   \path (N1_2) edge [red, line width=1.5 pt] (N1_4);
   \path (N1_2) edge [red, line width=1.5 pt] (N1_6);
   \path (N1_2) edge [red, line width=1 pt] (N1_7);
   
   \path (N1_3) edge [red, line width=0.5 pt] (N1_5);
   \path (N1_3) edge [red, line width=1.5 pt] (N1_9);
   
   \path [red, draw, line width = 1.5 pt] (center) (N1_4) arc (7 *360/\nrnodes: 8 *360/\nrnodes:\radius) ;
   \path (N1_4) edge [red, line width=1.5 pt] (N1_8);
   \path (N1_4) edge [red, line width=0.5 pt] (N1_9);
   
   \path [red, draw, line width = 0.5 pt] (center) (N1_5) arc (8 *360/\nrnodes: 9 *360/\nrnodes:\radius) ;
   \path (N1_5) edge [red, line width=0.5 pt] (N1_7);
   
   \path [red, draw, line width = 0.5 pt] (center) (N1_6) arc (9 *360/\nrnodes: 10 *360/\nrnodes:\radius) ;
   \path (N1_6) edge [red, line width=1.5 pt] (N1_9);
   
      \path [red, draw, line width = 1 pt] (center) (N1_7) arc (10 *360/\nrnodes: 11 *360/\nrnodes:\radius) ;
   \path (N1_7) edge [red, line width=0.5 pt] (N1_9);
   
   \path [red, draw, line width = 1.5 pt] (center) (N1_8) arc (11 *360/\nrnodes: 12 *360/\nrnodes:\radius) ;
  
   \coordinate (center) at (3,-4);
   \def \radius  {1.6}
   \def \nrnodes {8}
   
   \path [draw] (center) ++ ( -1 *360/\nrnodes:\radius) node [node2] (N2_1) {};
   \path [draw] (center) ++ ( 0 *360/\nrnodes:\radius) node [node2] (N2_2) {};
   \path [draw] (center) ++ ( 1 *360/\nrnodes:\radius) node [node2] (N2_3) {};
   \path [draw] (center) ++ ( 2 *360/\nrnodes:\radius) node [node2] (N2_4) {};
   \path [draw] (center) ++ ( 3 *360/\nrnodes:\radius) node [node2] (N2_5) {};
   \path [draw] (center) ++ ( 4 *360/\nrnodes:\radius) node [node2] (N2_6) {};
   \path [draw] (center) ++ ( 5 *360/\nrnodes:\radius) node [node2] (N2_7) {};
   \path [draw] (center) ++ ( 6 *360/\nrnodes:\radius) node [node2] (N2_8) {};
   
   \path [green, draw, line width = 1 pt] (center) (N2_1) arc (-1 *360/\nrnodes: 0 *360/\nrnodes:\radius) ;
   \path (N2_1) edge [green, line width=1 pt] (N2_5);
   \path (N2_1) edge [green, line width=0.5 pt] (N2_7);
   \path [green, draw, line width = 1.5 pt] (center) (N2_8) arc (6 *360/\nrnodes: 7 *360/\nrnodes:\radius) ;
    
   \path [green, draw, line width = 1 pt] (center) (N2_2) arc (0 *360/\nrnodes: 1 *360/\nrnodes:\radius) ;
   \path (N2_2) edge [green, line width=1 pt] (N2_5);
   \path (N2_2) edge [green, line width=1.5 pt] (N2_6);
   \path (N2_2) edge [green, line width=1 pt] (N2_7);
   \path (N2_2) edge [green, line width=1 pt] (N2_8);
   
   \path [green, draw, line width = 1.5 pt] (center) (N2_3) arc (1 *360/\nrnodes: 2 *360/\nrnodes:\radius) ;
   \path (N2_3) edge [green, line width=0.5 pt] (N2_8);
   
   \path [green, draw, line width = 1.5 pt] (center) (N2_4) arc (2 *360/\nrnodes: 3 *360/\nrnodes:\radius) ;
   \path (N2_4) edge [green, line width=1.5 pt] (N2_7);
   \path (N2_4) edge [green, line width=1 pt] (N2_8);
   
   \path [green, draw, line width = 0.5 pt] (center) (N2_5) arc (3 *360/\nrnodes: 4 *360/\nrnodes:\radius) ;
   \path (N2_5) edge [green, line width=1 pt] (N2_7);
   
   \path [green, draw, line width = 1 pt] (center) (N2_6) arc (4 *360/\nrnodes: 5 *360/\nrnodes:\radius) ;
   
      \path [green, draw, line width = 0.5 pt] (center) (N2_7) arc (5 *360/\nrnodes: 6 *360/\nrnodes:\radius) ;
   
   \coordinate (center) at (8,-1);
   \def \radius  {2.3}
   \def \nrnodes {10}
   
   \path [draw] (center) ++ ( 1 *360/\nrnodes:\radius) node [node3] (N3_1) {};
   \path [draw] (center) ++ ( 2 *360/\nrnodes:\radius) node [node3] (N3_2) {};
   \path [draw] (center) ++ ( 3 *360/\nrnodes:\radius) node [node3] (N3_3) {};
   \path [draw] (center) ++ ( 4 *360/\nrnodes:\radius) node [node3] (N3_4) {};
   \path [draw] (center) ++ ( 5 *360/\nrnodes:\radius) node [node3] (N3_5) {};
   \path [draw] (center) ++ ( 6 *360/\nrnodes:\radius) node [node3] (N3_6) {};
   \path [draw] (center) ++ ( 7 *360/\nrnodes:\radius) node [node3] (N3_7) {};
   \path [draw] (center) ++ ( 8 *360/\nrnodes:\radius) node [node3] (N3_8) {};
   \path [draw] (center) ++ ( 9 *360/\nrnodes:\radius) node [node3] (N3_9) {};
   \path [draw] (center) ++ ( 10 *360/\nrnodes:\radius) node [node3] (N3_10) {};
   
   \path (N3_1) edge [blue, line width=1 pt] (N3_3);
   \path (N3_1) edge [blue, line width=1 pt] (N3_4);
   \path (N3_1) edge [blue, line width=1.5 pt] (N3_5);
   \path (N3_1) edge [blue, line width=1.5 pt] (N3_7);
   \path (N3_1) edge [blue, line width=0.5 pt] (N3_8);
   \path (N3_1) edge [blue, line width=0.5 pt] (N3_9);
   \path [blue, draw, line width = 1 pt] (center) (N3_10) arc (10 *360/\nrnodes: 11 *360/\nrnodes:\radius) ;
   
   \path (N3_2) edge [blue, line width=1 pt] (N3_4);
   \path (N3_2) edge [blue, line width=0.5 pt] (N3_7);
   \path (N3_2) edge [blue, line width=0.5 pt] (N3_9);
   
   \path [blue, draw, line width = 1.5 pt] (center) (N3_3) arc (3 *360/\nrnodes: 4 *360/\nrnodes:\radius) ;
   \path (N3_3) edge [blue, line width=1.5 pt] (N3_6);
   \path (N3_3) edge [blue, line width=1 pt] (N3_8);
   \path (N3_3) edge [blue, line width=1 pt] (N3_9);
   \path (N3_3) edge [blue, line width=1.5 pt] (N3_10);
   
   \path (N3_4) edge [blue, line width=1 pt] (N3_8);
   \path (N3_4) edge [blue, line width=0.5 pt] (N3_10);
   
   \path (N3_5) edge [blue, line width=1.5 pt] (N3_7);
   \path (N3_5) edge [blue, line width=0.5 pt] (N3_8);
   \path (N3_5) edge [blue, line width=0.5 pt] (N3_10);
   
   \path [blue, draw, line width = 1 pt] (center) (N3_6) arc (6 *360/\nrnodes: 7 *360/\nrnodes:\radius) ;
   \path (N3_6) edge [blue, line width=1 pt] (N3_9);
   \path (N3_6) edge [blue, line width=0.5 pt] (N3_10);
   
   \path [blue, draw, line width = 1.5 pt] (center) (N3_8) arc (8 *360/\nrnodes: 9 *360/\nrnodes:\radius) ;
   
   \path [blue, draw, line width = 1.5 pt] (center) (N3_9) arc (9 *360/\nrnodes: 10 *360/\nrnodes:\radius) ;
   

   \path (N3_4) edge [line width=0.5 pt] (N1_5);
   \path (N3_6) edge [line width=0.5 pt] (N2_3);
   \path (N3_5) edge [line width=0.5 pt] (N1_6);
   
   \end{tikzpicture}
  \caption{}
  \label{fig_synthetic_network_1}
\end{subfigure}%
\begin{subfigure}{.65\textwidth}
  \centering
    	\input{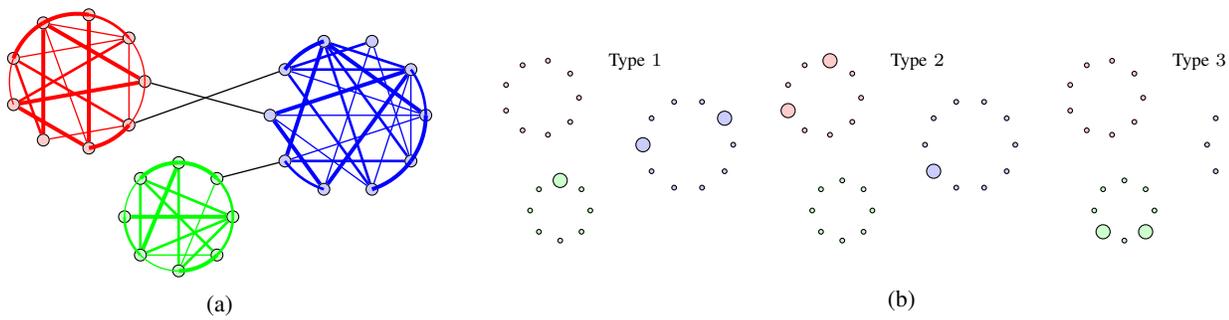}
  \caption{}
  	\label{fig_synthetic_network_2}
\end{subfigure}

\caption{(a) The three-cluster network on which signals to be classified are defined. The width of the links is proportional to the weights of the corresponding edges. (b) Sample signals for the three types considered. Type 1 signals have stronger presence in the blue cluster, type 2 in the red, and type 3 in the green cluster.}
\label{fig_synthetic_network}
\end{figure*}

In contrast to Theorem \ref{thm_spsnorm_stability}, the bound in \eqref{eqn_thm_diffnorm_stability} contains higher order terms that depend on the magnitude of the perturbation. Hence, since the other terms of the bound in \eqref{eqn_thm_diffnorm_stability} tend to zero super linearly, we may divide \eqref{eqn_thm_diffnorm_stability} by $\epsilon \|L\|_p$ and compute the limit as the perturbation vanishes
\begin{align}\label{eqn_thm_diffnorm_stability_limit}
      \lim_{\epsilon \rightarrow 0} \frac{\left| d_{\diff}^{L'}(s,r) - d_{\diff}^{L}(s,r)\right|}
            {\epsilon \|L\|_p} \le 2 \gamma,
\end{align}
which implies that for small perturbations the difference in diffusion distances grows linearly.

When constructing the underlying graph to compare signals in a real-world application, noisy information can be introduced. This means that the similarity weight between two nodes in the underlying graph contains inherent error. Theorems \ref{thm_spsnorm_stability} and \ref{thm_diffnorm_stability} show that the superposition and diffusion distances are impervious to these minor perturbations.

In order to illustrate the stability results presented, consider again the underlying network in Figure \ref{fig_network_example_diffusion}. We perturb this network by multiplying every edge weight -- originally equal to 1 -- by a random number uniformly picked from $[0.95, 1.05]$ and then compute the diffusion and superposition distances between vectors $r$ and $g$ with the perturbed graph as underlying network. For these illustrations we pick the input norm to be $\ell_2$ and observe that $\gamma = 1$ given the definitions of $r$ and $g$. In Figure \ref{fig_stability_hist} we plot histograms of the absolute value of the difference in the distances when using the original and the perturbed graphs as underlying networks normalized by the norm of the perturbation for 1000 repetitions of the experiment. From \eqref{eqn_thm_spsnorm_stability} we know that this value should be less than 2 for the superposition distance and from \eqref{eqn_thm_diffnorm_stability_limit} we know this should also be the case for the diffusion distance for vanishing perturbations. Indeed, as can be seen from Figure \ref{fig_stability_hist}, all perturbations are below the threshold of 2 by a considerable margin. This stability property is essential for the practical utility of the diffusion and superposition distances as seen in the next section.

\begin{remark}\label{rem_stability_norms}\normalfont
In Theorems \ref{thm_spsnorm_stability} and \ref{thm_diffnorm_stability} we focus our analysis on the input norms $\| \cdot \|_p$ for $p \in \{1, 2, \infty \}$ because these norms lead to the simple bounds in \eqref{eqn_thm_spsnorm_stability} and \eqref{eqn_thm_diffnorm_stability}. The simplicity of these bounds is derived from the fact that $\| e^{-Lt} \|_p \leq 1$ and $\| (I + L)^{-1} \|_p \leq 1$ for the values of $p$ previously mentioned. For other matrix norms satisfying \eqref{eqn_definition_submultiplicativity} and \eqref{eqn_definition_compatibility}, including all induced matrix norms, the equivalence of norms guarantees that bounds analogous to those in \eqref{eqn_thm_spsnorm_stability} and \eqref{eqn_thm_diffnorm_stability} must exist with more complex constant terms.
\end{remark}

%
\begin{figure}[t]
    	\centering
	
\begin{subfigure}{.25\textwidth}
  \centering
    	\includegraphics[width = 1\textwidth, keepaspectratio]{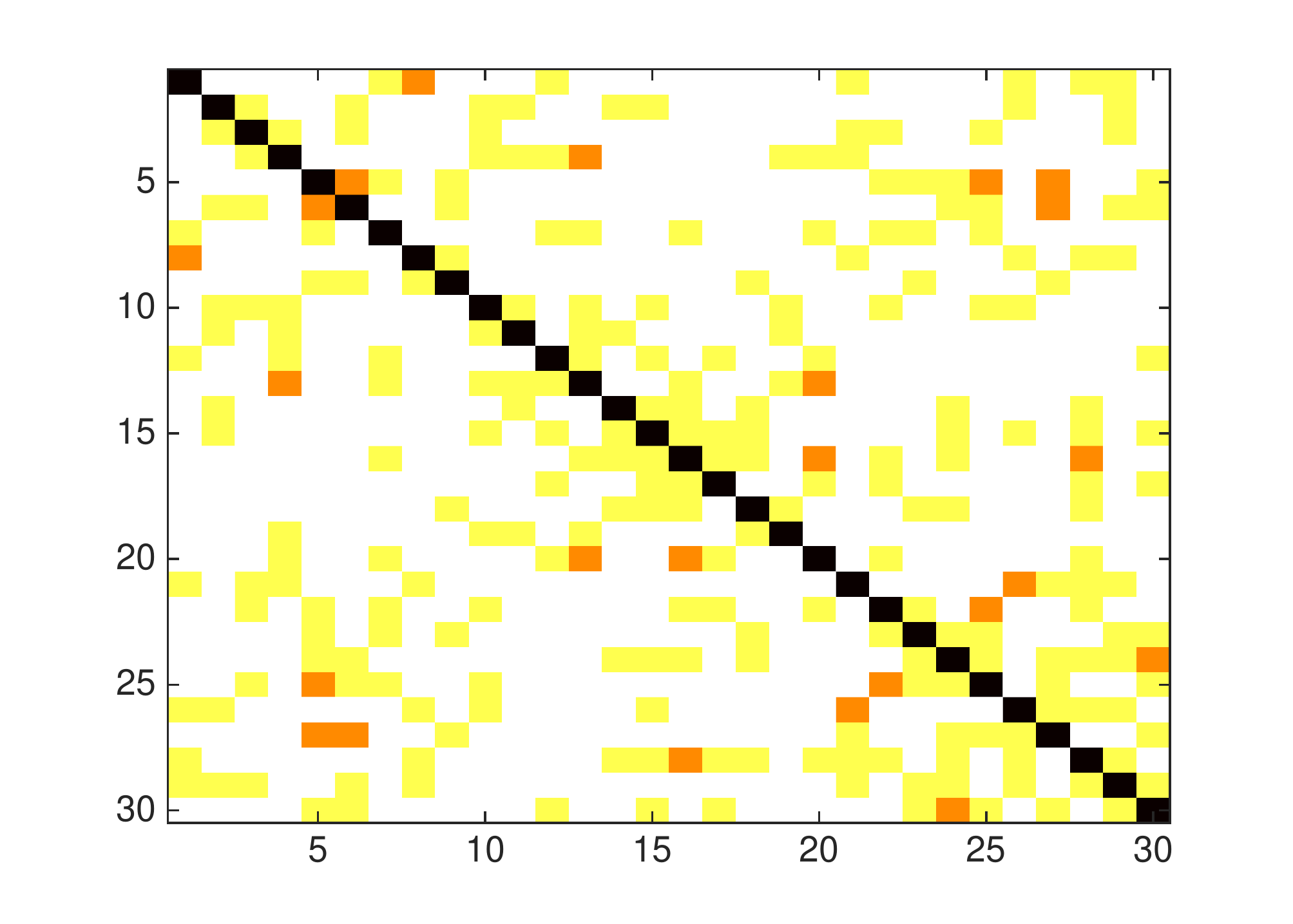}
  \caption{$\ell_2$ heat map}
  \label{fig_synthetic_network_results_1}
\end{subfigure}%
\begin{subfigure}{.25\textwidth}
  \centering
    	\includegraphics[width = 1\textwidth, keepaspectratio]{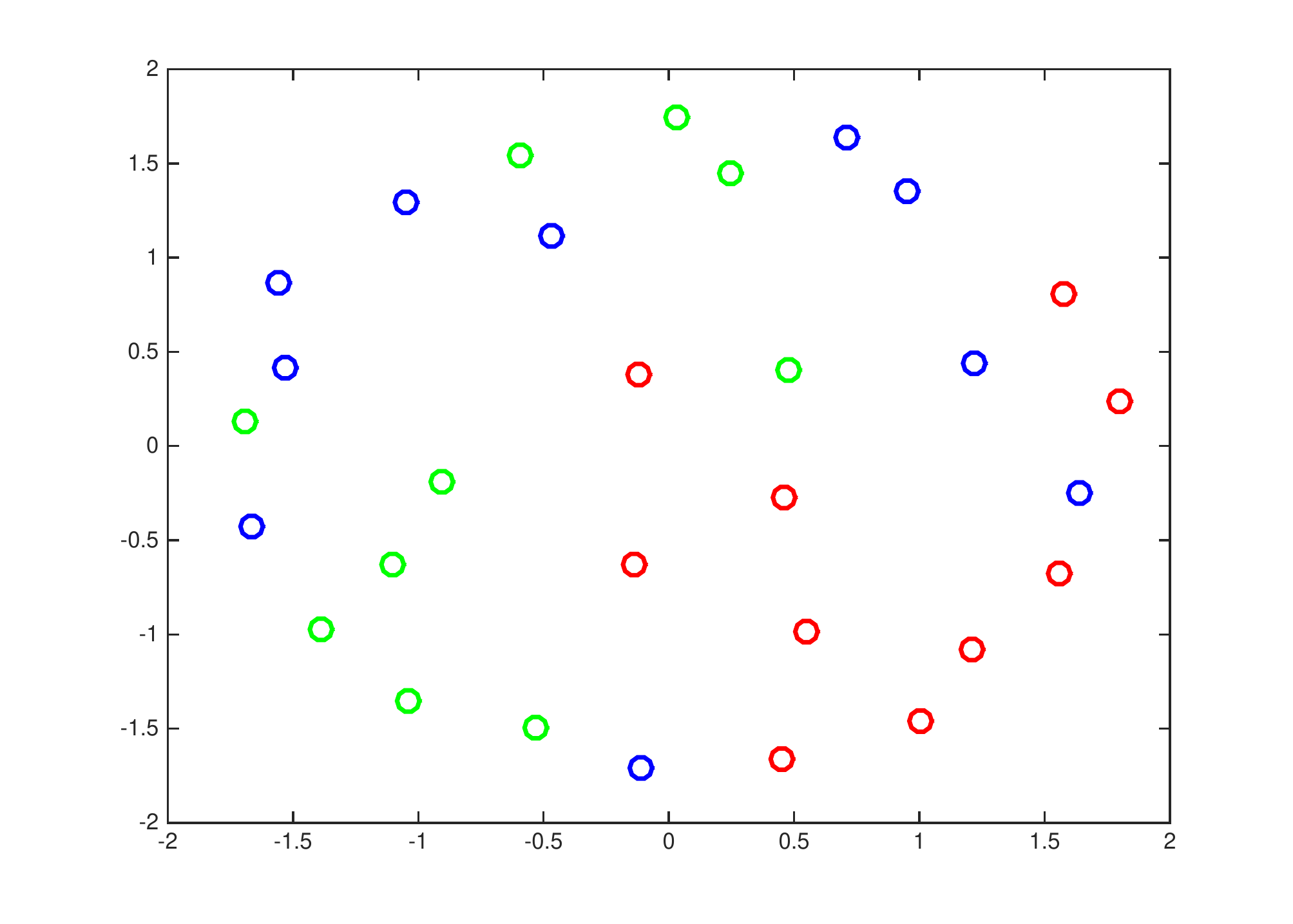}
  \caption{MDS for $\ell_2$}
  	\label{fig_synthetic_network_results_2}
\end{subfigure}\\
\begin{subfigure}{.25\textwidth}
  \centering
    	\includegraphics[width = 1\textwidth, keepaspectratio]{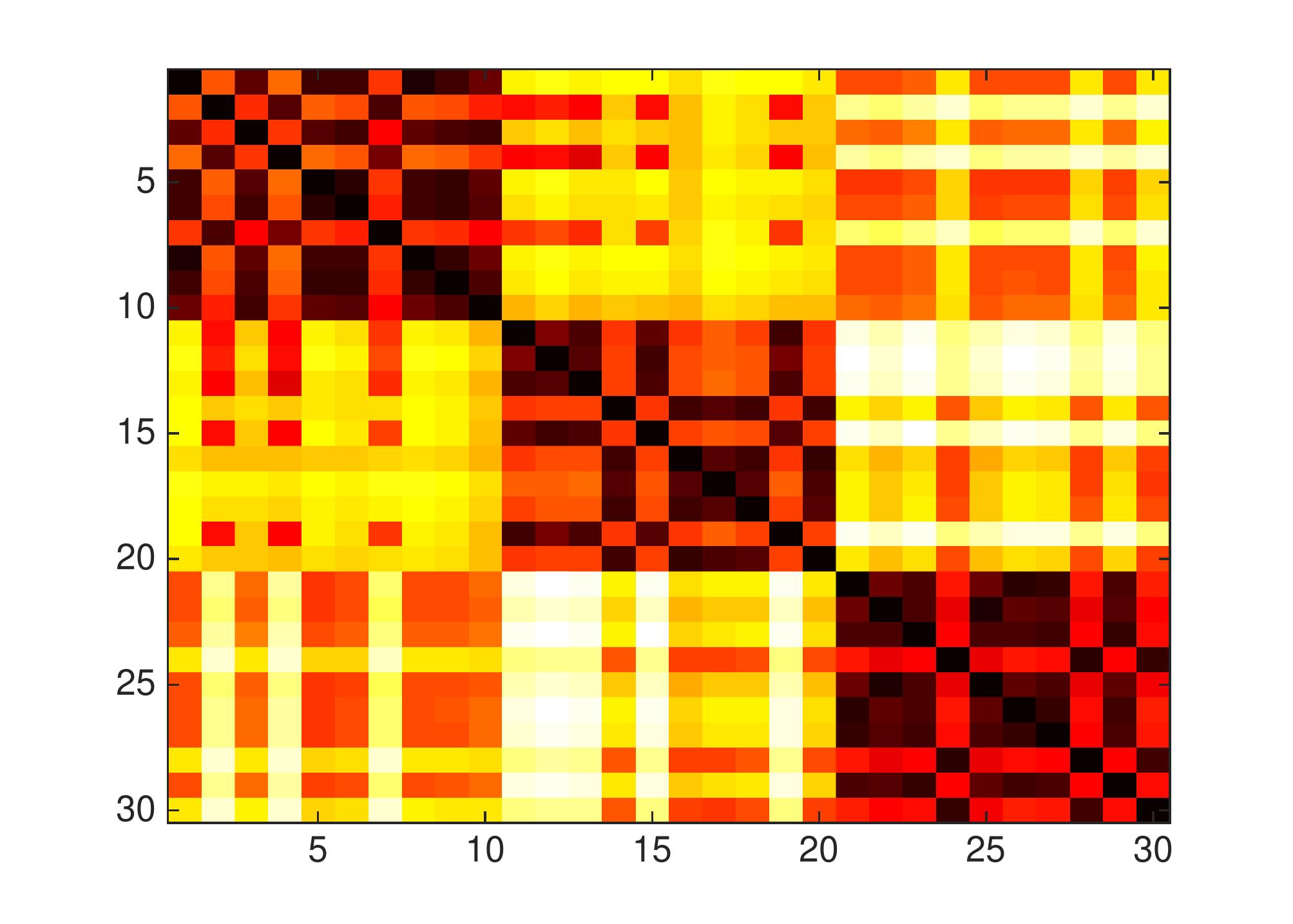}
  \caption{Diffusion heat map}
  \label{fig_synthetic_network_results_3}
\end{subfigure}%
\begin{subfigure}{.25\textwidth}
  \centering
    	\includegraphics[width = 1\textwidth, keepaspectratio]{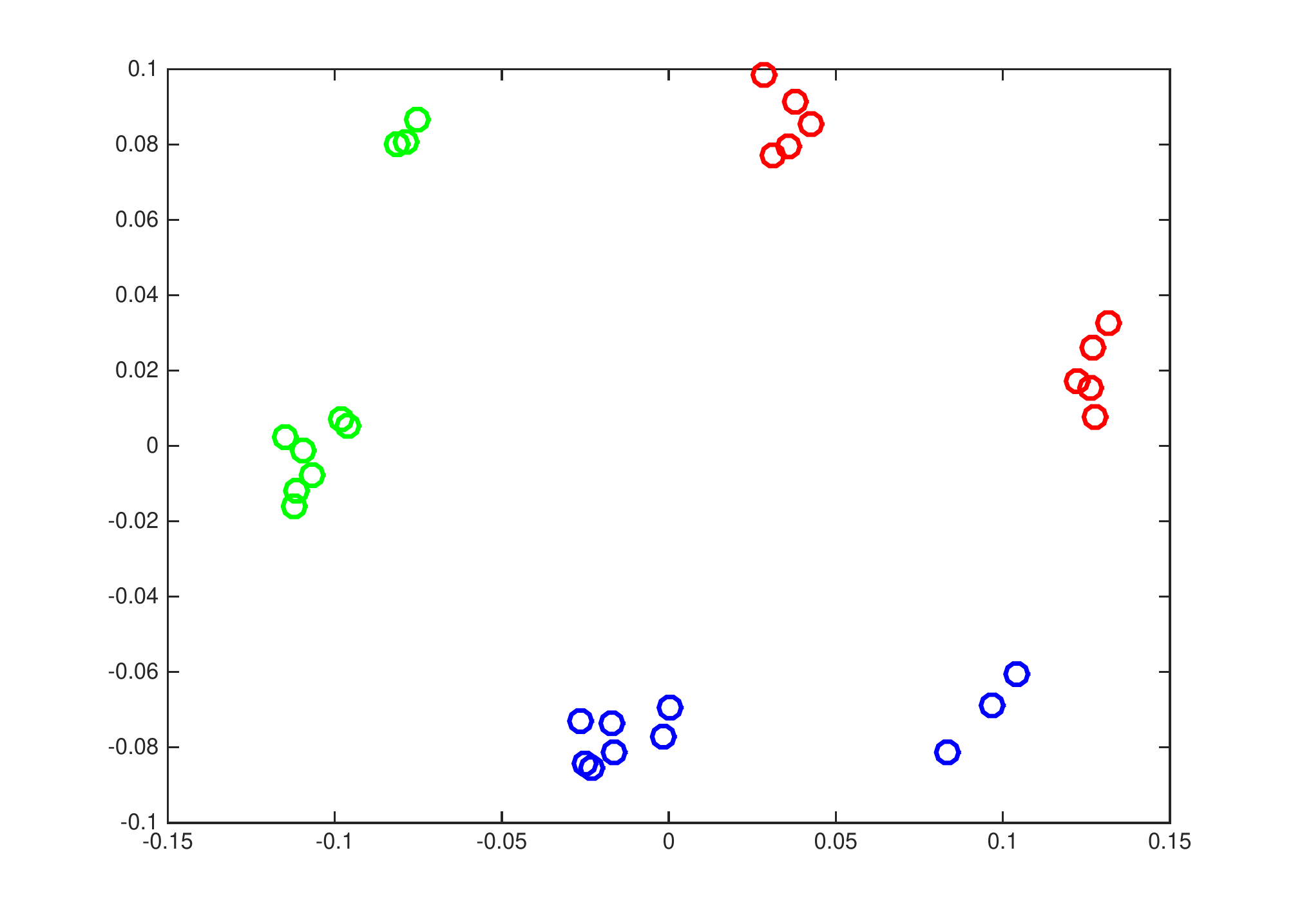}
  \caption{MDS for diffusion}
  	\label{fig_synthetic_network_results_4}
\end{subfigure}\\
\begin{subfigure}{.25\textwidth}
  \centering
    	\includegraphics[width = 1\textwidth, keepaspectratio]{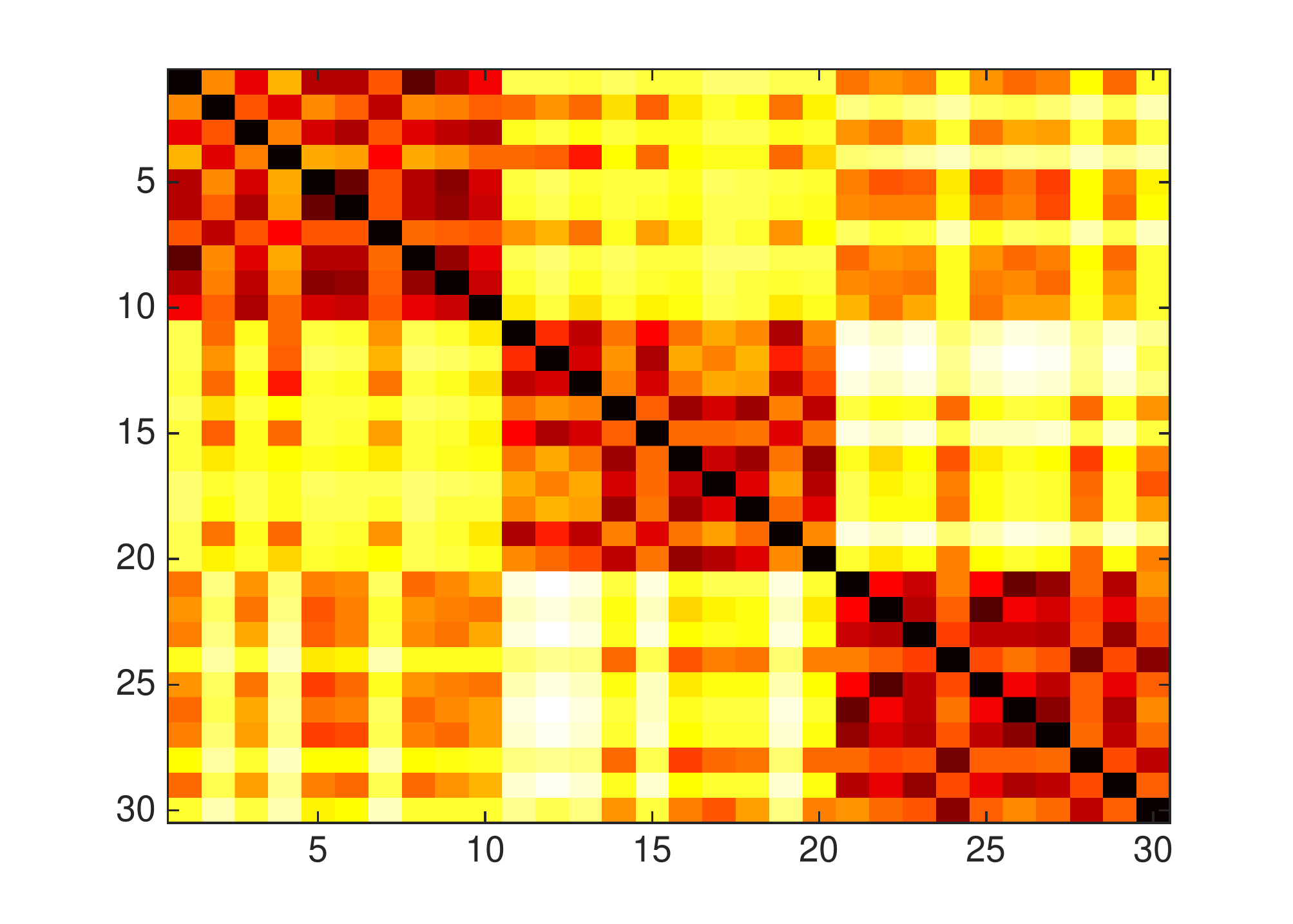}
  \caption{Superposition heat map}
  \label{fig_synthetic_network_results_5}
\end{subfigure}%
\begin{subfigure}{.25\textwidth}
  \centering
    	\includegraphics[width = 1\textwidth, keepaspectratio]{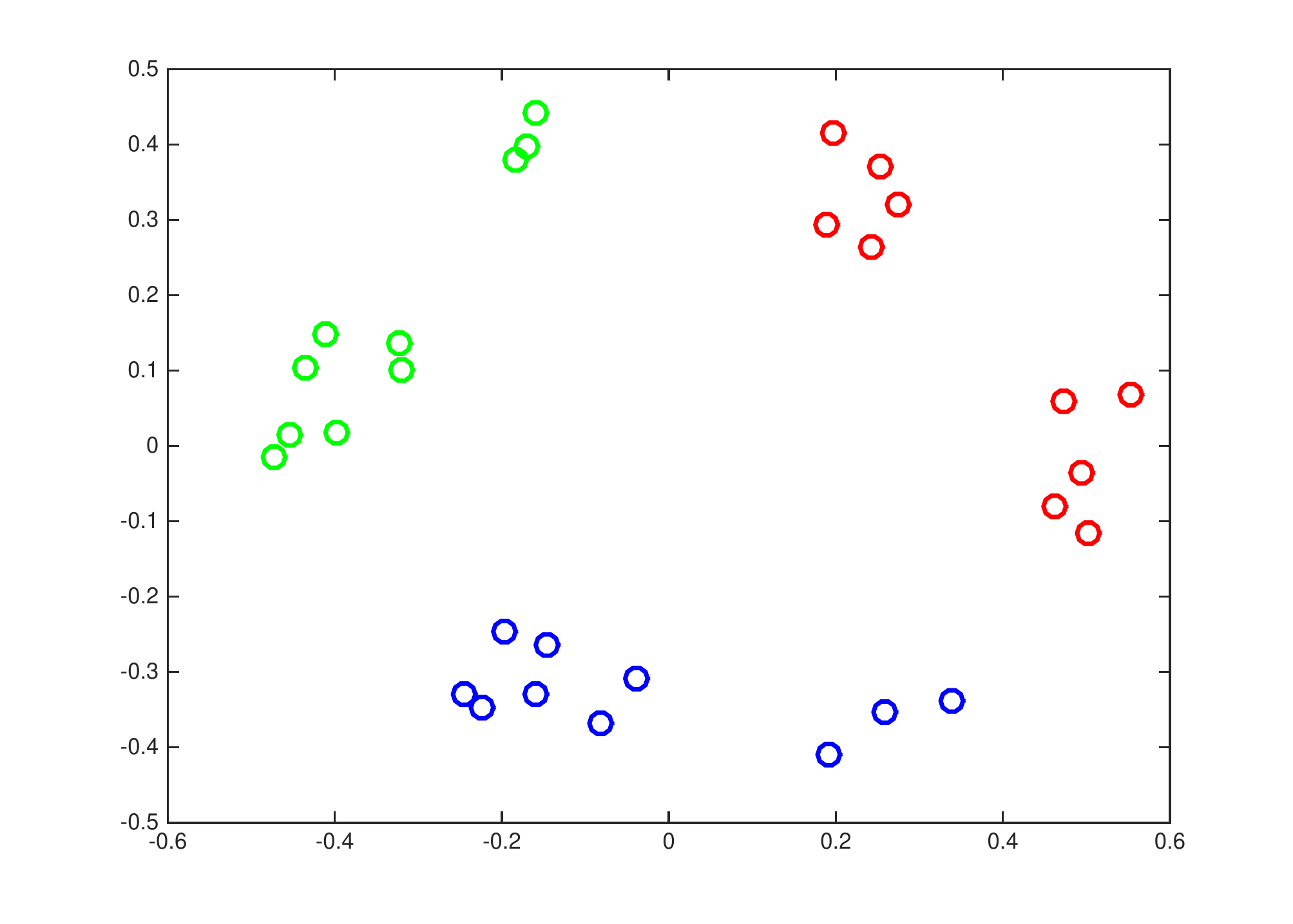}
  \caption{MDS for superposition}
  	\label{fig_synthetic_network_results_6}
\end{subfigure}

\caption{Heat maps (left) and 2 dimensional MDS representations (right) for the metric spaces generated by the $\ell_2$ (top), diffusion (middle) and superposition (bottom) distances. The diffusion and superposition metrics perfectly classify the signals into the three types while $\ell_2$ does not reveal any clear classification.}
\label{fig_synthetic_network_results}
\end{figure}



\section{Applications}\label{sec_applications}

We illustrate the advantages of the superposition and diffusion distances developed in Sections \ref{sec_superposition_distance} and \ref{sec_diffusion_distance} respectively through numerical experiments in both synthetic (Section \ref{sec_three_clusters}) and real-world data (Section \ref{sec_cancer_histology}).

%
\subsection{Classification of synthetic signals on networks}\label{sec_three_clusters}

The diffusion and superposition distances lead to better classification of signals on networks compared to traditional vector distances such as the Euclidean $\ell_2$ metric. Consider the network presented in Figure \ref{fig_synthetic_network_1} containing three clusters -- blue, red, and green -- where nodes within each cluster are highly connected and there exist few connections between nodes in different clusters. This network was generated randomly, where an undirected edge between a pair of nodes in the same cluster is formed with probability $0.4$ and its weight is picked uniformly between $1$ and $3$. In addition, three edges were added with weight $1$ between random pairs of nodes in different clusters. We consider three types of signals on this network. The strength of all signals is equal to $1$ on three nodes in the network and $0$ on the remaining ones. Among the three nodes with value $1$ for the first type of signals, two of them are randomly selected from the blue cluster and the remaining one is randomly chosen from the other clusters. Similarly, for the second type of signals, exactly two out of the three nodes with positive value belong to the red cluster and the remaining one is chosen randomly between the blue and green clusters. Finally, the third type of signal has two positive values on the green cluster and the third value randomly chosen from the rest of the network. Sample signals for each type are illustrated in Figure \ref{fig_synthetic_network_2} where positive signal values are denoted by larger nodes.

\begin{figure*}[t]
\centering

\begin{subfigure}{.5\textwidth}
  \centering
    \includegraphics[width=1 \textwidth, height=0.75 \textwidth]{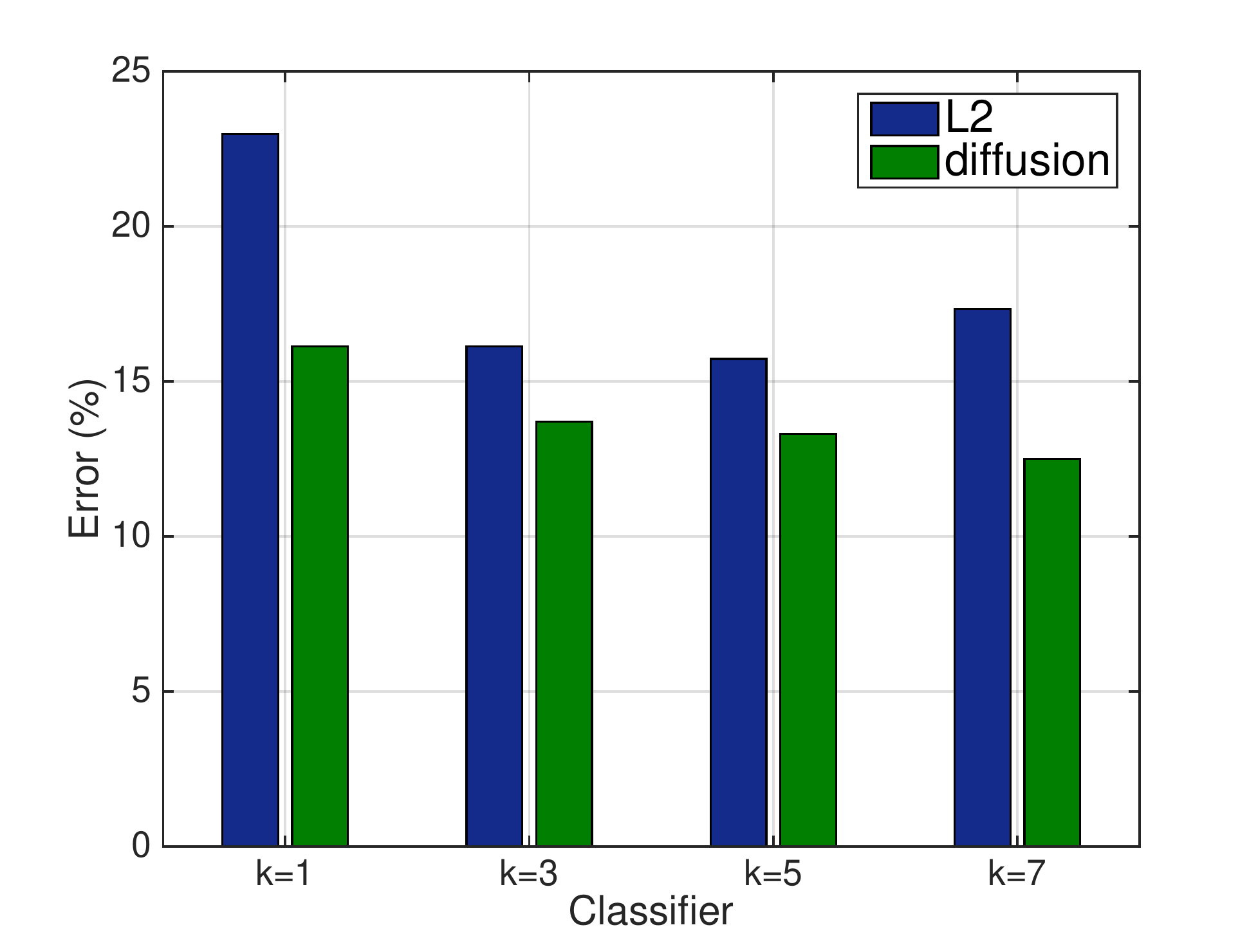}
  \caption{}
\label{fig_cancer_classification_1}
\end{subfigure}%
\begin{subfigure}{.5\textwidth}
  \centering
    \includegraphics[width=1 \textwidth, height=0.75 \textwidth]{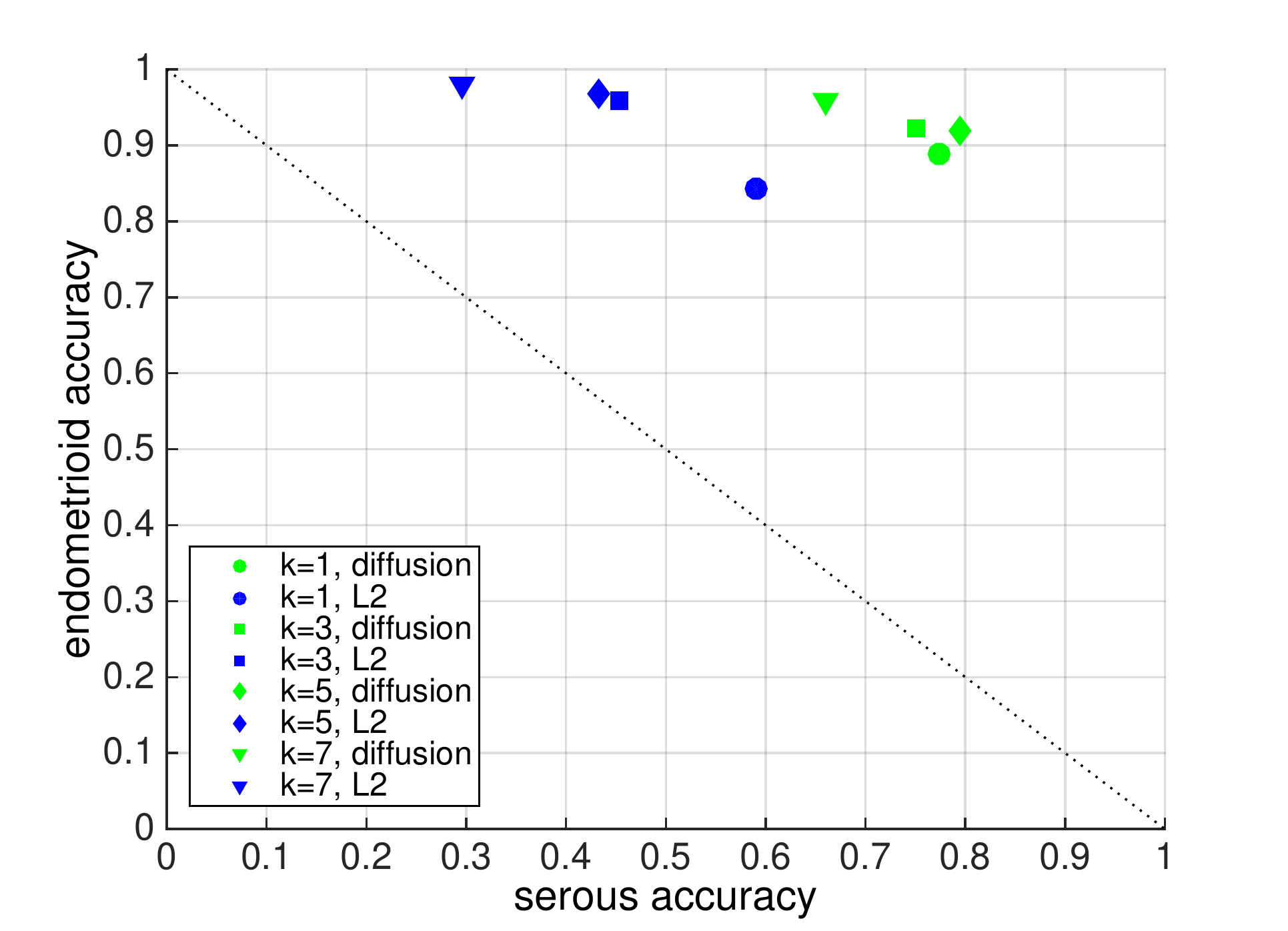}
  \caption{}
\label{fig_cancer_classification_2}
\end{subfigure}

\caption{Histology classification of ovarian cancer patients based on $k$ nearest neighbors with respect to the $\ell_2$ and diffusion distances of their genetic profile. (a) Blue bars denote the error when patients are classified using the $\ell_2$ distance while the green bars denote the error when diffusion distance is used for different k-NN classifiers. The diffusion distance reduces the classification error consistently across classifiers. (b) Accuracy of serous subtype vs. endometrioid subtype. Classifiers using diffusion (green) are closer to the top right corner, i.e. perfect classification, than those using the $\ell_2$ distance (blue).}
\label{fig_cancer_classification}
\end{figure*}

We generate ten signals of each type and measure the distance between them with the superposition, diffusion, and $\ell_2$ metrics. For the superposition and diffusion metrics we use $\ell_2$ as input norm and $\alpha=1$. The use of each metric generates a different metric space with the thirty signals as the common underlying set of points. In order to illustrate these higher dimensional spaces, in Figure \ref{fig_synthetic_network_results} (left) we present heat maps of the distance functions, where darker colors represent closer signals. It is clear that for the diffusion and superposition distances, three blocks containing ten points each appear along the diagonal in exact correspondence with the three types of signals. In contrast, the heat map corresponding to the $\ell_2$ metric does not present any clear structure. To further illustrate these implications, in Figure \ref{fig_synthetic_network_results} (right) we present 2D multi dimensional scaling (MDS) \cite{CoxCox08} representations of the three metric spaces. The points corresponding to type 1 signals are represented as blue circles, type 2 as red circles, and type 3 as green circles. The MDS representations for diffusion and superposition are fundamentally different from the one obtained for $\ell_2$. For the latter, the circles of different colors are spread almost randomly on the plane, with no clear clustering structure. For diffusion and superposition, in contrast, signals of different colors are clearly separated so that any clustering method is able to recover the original signal type.

%
\subsection{Ovarian cancer histology classification}\label{sec_cancer_histology}

We demonstrate that the diffusion distance can provide a better classification of histology subtypes for ovarian cancer patients than the traditional $\ell_2$ metric. To do this, we consider 240 patients diagnosed with ovarian cancer corresponding to two different histology subtypes \cite{Hofree13}: serous and endometrioid. Our objective is to recover the histology subtypes from patients' genetic profiles. 

For each patient $i$, her genetic profile consists of a binary vector $v_i \in \{0, 1\}^{2458}$ where, for each of the 2458 genes studied, $v_i$ contains a 1 in position $k$ if patient $i$ presents a mutation in gene $k$ and 0 otherwise. One way of building a metric in the space of 240 patients is by quantifying the distance between patients $i$ and $j$ as the $\ell_2$ distance between their genetic profiles,
\begin{align}\label{eqn_L2_genetic_distance}
d_{\ell_2}(i,j) = \|v_i - v_j \|_2.
\end{align}
In this approach, every gene is considered orthogonal to each other and compared separately across patients. An alternative approach is to take into account the relational information across genes when comparing patients. In order to do so, we apply the diffusion distance on an underlying gene-to-gene network built based on publicly available data \cite{PATHWAY}. In order to build this network, we first extract the pairwise gene-gene interactions from \cite{PATHWAY} using the \emph{NCI\_Nature} database. After normalization, every edge weight is contained between 0 and 1, which we interpret as a probability of interaction between genes. We assign to each path the probability obtained by multiplying the probabilities in the edges that form the path. For every pair of genes in the network, we compute a similarity value between them corresponding to the maximum probability achievable by a path that links both genes. Finally, we apply normalization and thresholding operations to obtain the gene-to-gene network that we use in our experiments. Observe that the gene-to-gene network contains accepted relations between genes in humans in general and is not patient dependent, hence, it defines a common underlying network for all subjects being compared. Thus, denoting as $L$ the Laplacian of the gene-to-gene network and using the $\ell_2$ as input norm we compute the diffusion distances between patients $i$ and $j$ as [cf. \eqref{eqn_definition_diffusion_distance_2}]
\begin{align}\label{eqn_diffusion_genetic_distance}
d^L_\diff(i,j) = \|(I + \alpha L) ^{-1} (v_i - v_j) \|_2,
\end{align}
where $\alpha$ was set to 15, however, results are robust to this particular choice. Given that in Section \ref{sec_three_clusters} we obtained similar performance between the diffusion and superposition distances, combined with the fact that the latter is computationally expensive, we do not implement the superposition distance in this data set.

In order to evaluate the classification power of both approaches -- $\ell_2$ and diffusion distance -- we perform $240$-fold cross validation for a $k$ nearest neighbors (k-NN) classifier. More precisely, for a particular patient, we look at the $k$ nearest patients as given by the metric being evaluated and assign to this patient the most common cancer histology among the $k$ nearest patients. We then compare the assigned histology with her real cancer histology and evaluate the accuracy of the classifier. Finally, we repeat this process for the 240 women considered and obtain a global classification accuracy of both approaches.

In Figure \ref{fig_cancer_classification_1} we show the reduction in histology classification error when using the diffusion distance \eqref{eqn_diffusion_genetic_distance} compared to using the $\ell_2$ distance \eqref{eqn_L2_genetic_distance} when comparing genetic profiles. 
The four groups of bars correspond to classifiers built using different numbers of neighbors $k \in \{1, 3, 5, 7\}$. Notice that the reduction in error is consistent across all classifiers analyzed with an average reduction of over 4\% in the error rates, unveiling the value of incorporating the network information in the classification process.

To further analyze the obtained results, in Figure \ref{fig_cancer_classification_2} we present the accuracy obtained for the serous subtype versus the accuracy obtained for the endometrioid subtype for different classifiers based on the diffusion (green) and $\ell_2$ (blue) distances. Points on the top right corner of the plot are ideal, obtaining perfect classification for both subtypes. When using diffusion, accuracies shift towards the ideal position since the accuracies for the serous subtypes increase by $20\%$ to $40\%$ whereas the accuracies for endometrioid subtypes decrease by less than $5\%$. Furthermore, among the 240 patients analyzed, there are 196 of them with endometrioid subtype and only 44 with serous subtype. Hence, a nearest neighbor classifier based on an uninformative distance would tend to have a high classification accuracy for the former but a low one for the latter. This is the case for the $\ell_2$ metric. The diffusion distance, in contrast, by exploiting the gene-to-gene interaction can overcome this limitation.


\section{Feature space transformation}\label{sec_feature_space_transformation}

The diffusion distance in \eqref{eqn_definition_diffusion_distance_2} between $r, s \in \reals^n$ can be interpreted as the input norm of the difference between two diffused vectors $r_\diff$ and $s_\diff$ also defined in $\reals^n$, i.e. $d^L_{\diff}(r, s)=\| r_\diff - s_\diff \|$ where
\begin{equation}\label{eqn_def_diffused_signals}
r_\diff = (I + \alpha L)^{-1} r,
\end{equation}
and similarly for $s_\diff$. Thus, diffusion can be seen as a transformation of the feature space for cases where there exists additional information about the relation between features. This relation is based on prior knowledge about the feature spaced instead of being data driven by particular observations. For example, for the genetic network in Section \ref{sec_cancer_histology} we have the additional information -- independent of the set of patients -- that there is interrelation between the function of some genes. Hence, we use these relations to define diffused mutations for each patient. However, apart from looking at the distance between the diffused signals -- as proposed in Section \ref{sec_diffusion_distance} and applied in Section \ref{sec_applications} -- we can analyze the image of each signal under this transformation.

\begin{figure}[t]
\begin{minipage}[h]{0.24\textwidth}
    	\centering
	\hspace{-0.2in}
    	\includegraphics[height=0.95\textwidth, width = 1.07\textwidth, keepaspectratio]{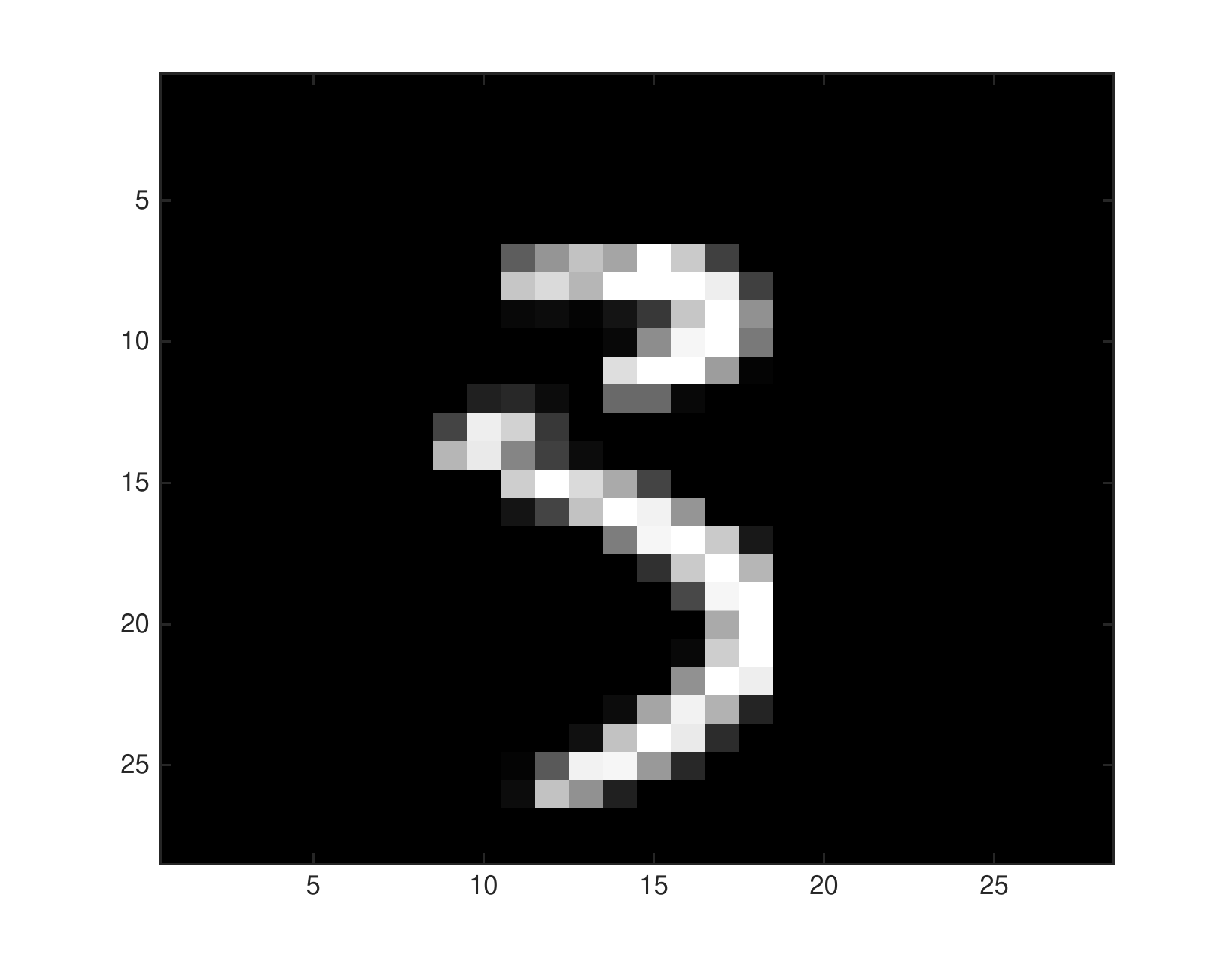}
\end{minipage}
\begin{minipage}[h]{0.24\textwidth}
    	\centering
    	\includegraphics[height=0.95\textwidth, width = 1.07\textwidth, keepaspectratio]{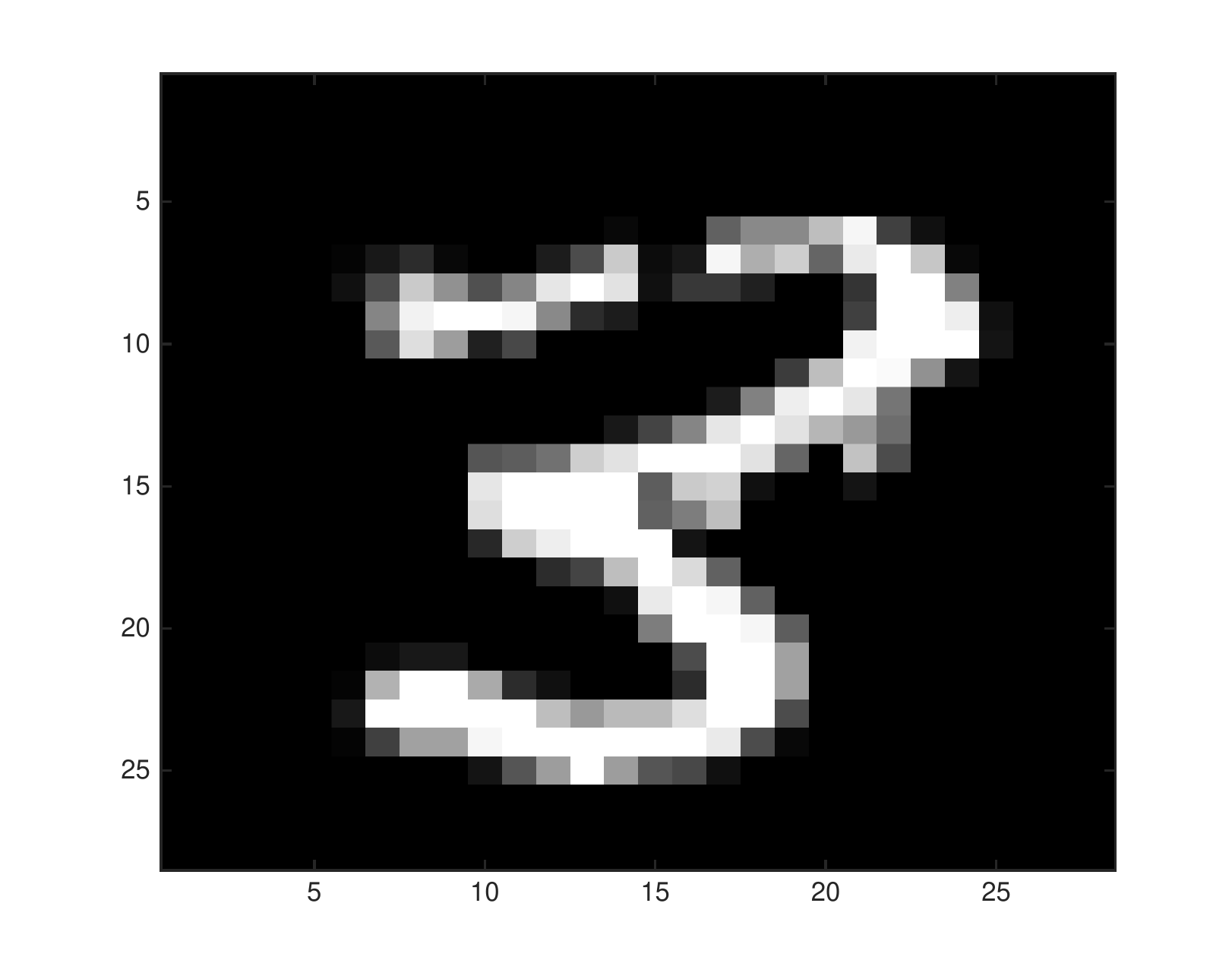}
\end{minipage}
\\
\begin{minipage}[h]{0.24\textwidth}
    	\centering
		\hspace{-0.2in}
    	\includegraphics[height=0.95\textwidth, width = 1.07\textwidth, keepaspectratio]{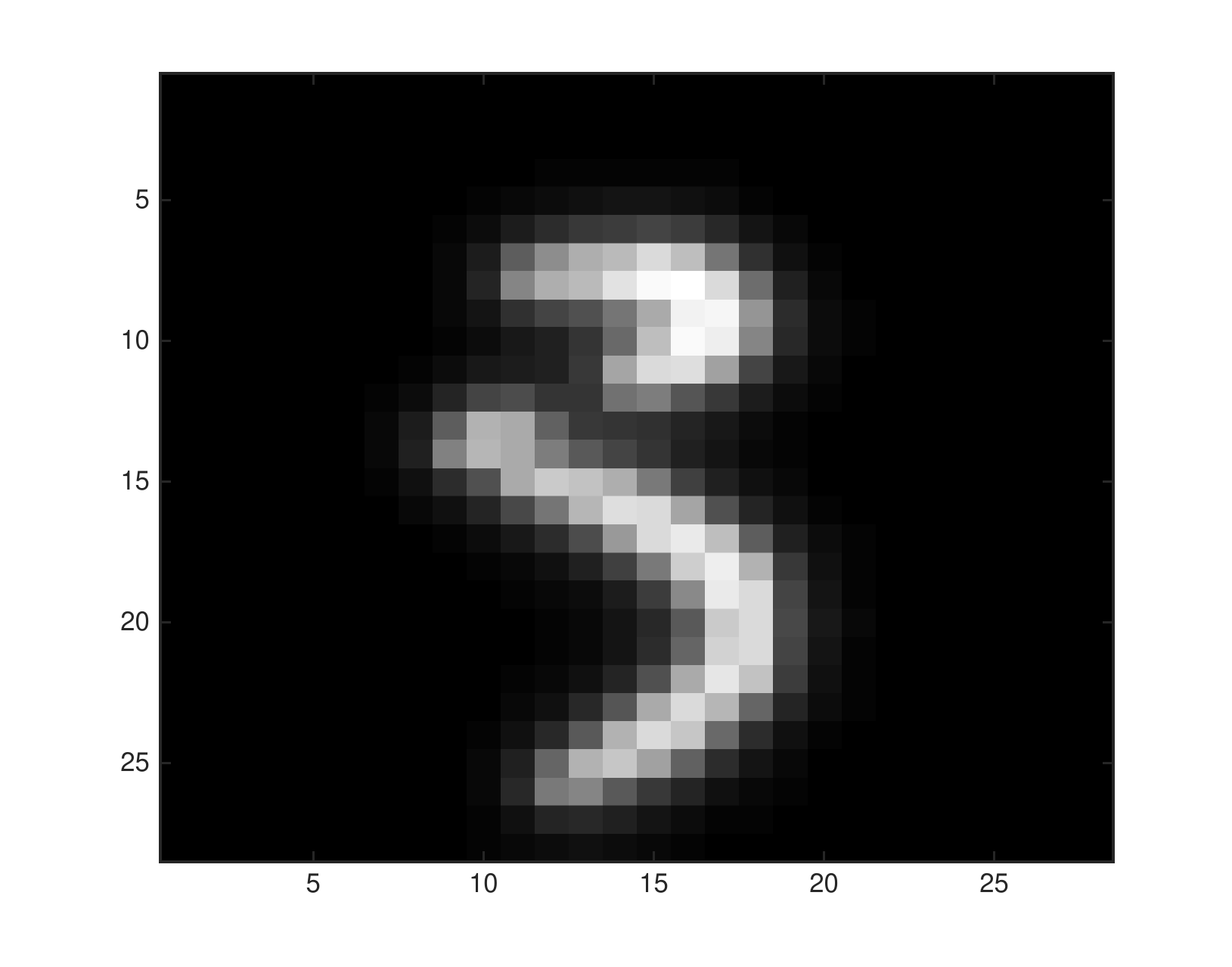}
\end{minipage}
\begin{minipage}[h]{0.24\textwidth}
    	\centering
    	\includegraphics[height=0.95\textwidth, width = 1.07\textwidth, keepaspectratio]{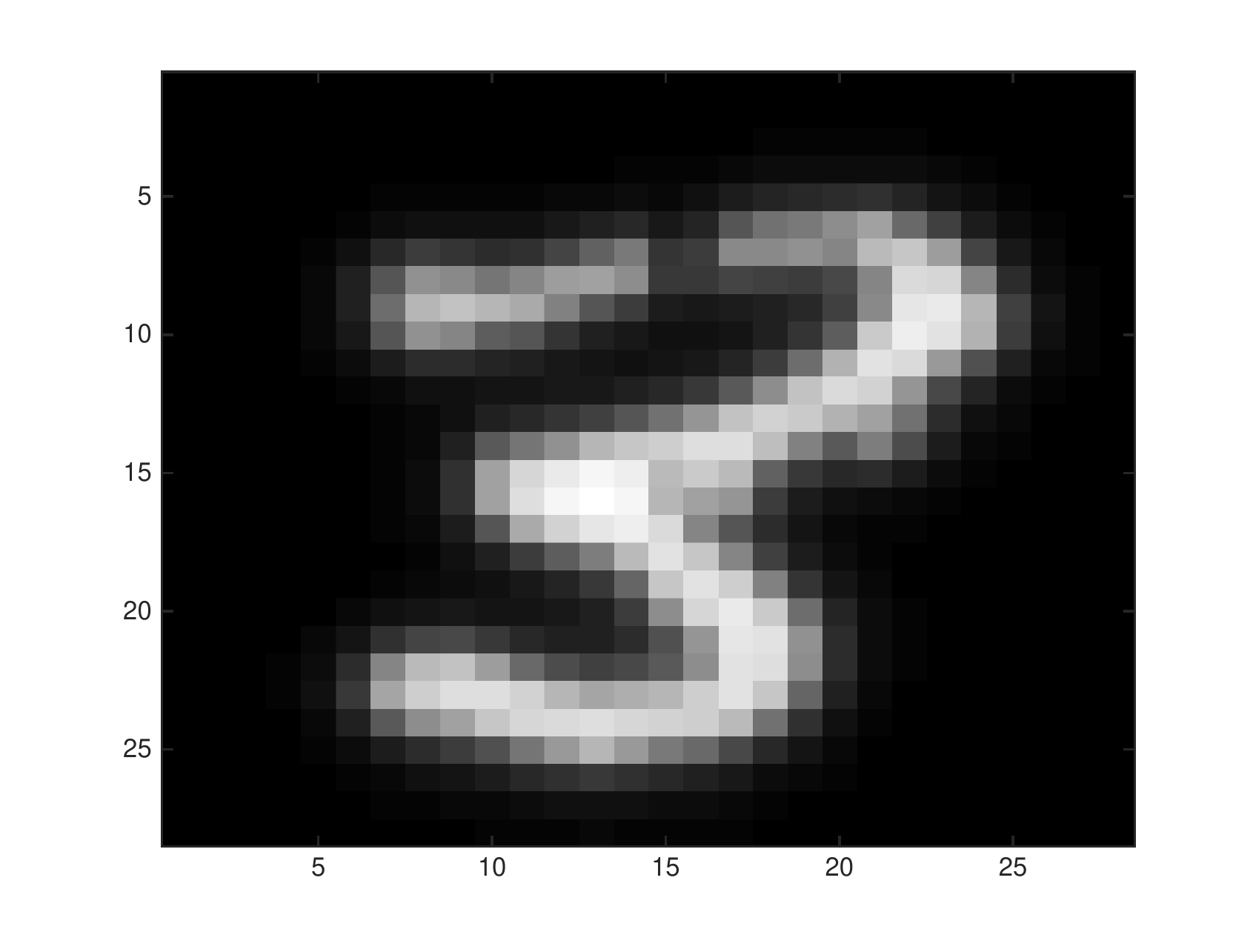}
\end{minipage}

\caption{Two samples of the handwritten digit 3 (top row) and their corresponding diffused versions (bottom row). The imperfections of the original handwritten samples are smoothened by diffusion making it easier to compare the diffused versions of the digits.}
\label{fig_number_3_comparison}
\end{figure}

As an illustration, consider the well-known MNIST handwritten digit database \cite{MNIST}. Each observation consists of a square gray-scaled image of a handwritten digit with 28 $\times$ 28 pixels. Consequently, we can think of each observation as a vector $x \in \reals^{784}$ where the value of each component corresponds to the intensity of the associated pixel. However, among these 784 features there are relations imposed by the lattice structure of the image. In particular, pixels found close in the image play a similar role in the specification of a handwritten digit. Thus, we build a lattice graph where each pixel is linked by an edge of unit weight to its contiguous pixels. If we denote by $L$ the Laplacian of the lattice graph built, we may use \eqref{eqn_def_diffused_signals} to obtain the diffused versions of different handwritten digits. In Figure \ref{fig_number_3_comparison} we present two different observations of the digit 3 as found in the MNIST database and after diffusion with $\alpha = 0.8$. From the figure, it is clear that diffusion smoothens imperfections of particular hand written instances, facilitating the comparison of diffused versions of the digits. E.g., the $\ell_2$ distance between the two original images is 10.13 while the average distance between any two digits taken at random from the database is 10.19. However, after diffusion, the $\ell_2$ distance between these two images is 6.41 while the average distance between any two diffused digits is 6.88, providing a better classification power. This motivates the use of diffusion as a preprocessing transformation for learning.

%

\begin{figure}[t]
  \centering
    \includegraphics[width=0.5 \textwidth]{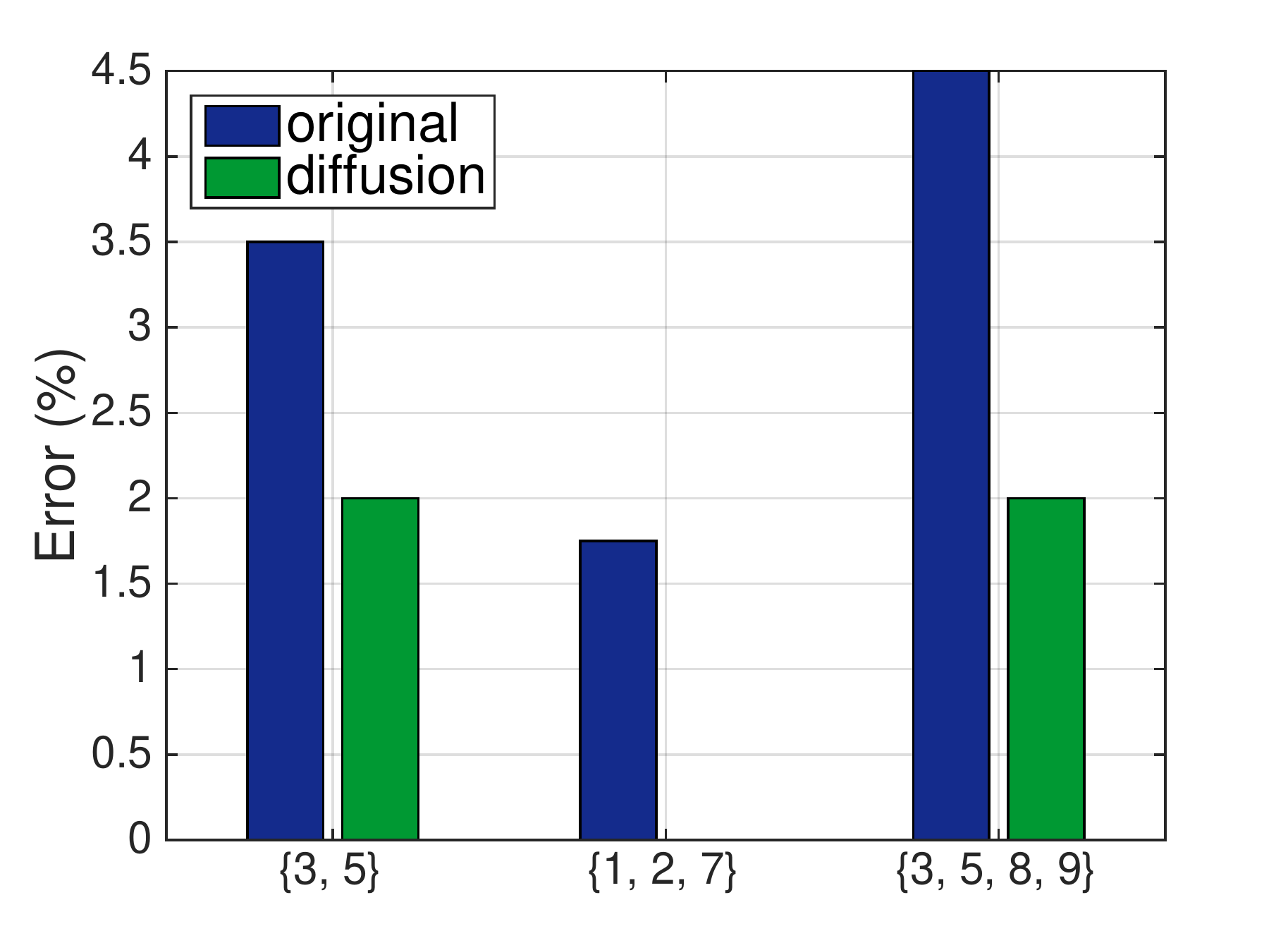}
\caption{Error rates for a two class, a three class and a four class classification of written digits given by a SVM trained in the original and the diffused data set. The error is reduced by diffusion in the three cases.}
\label{fig_number_classification_error}
\end{figure}

In this direction, we build two support vector machine (SVM) classifiers with a radial basis function kernel \cite{Scholkopf97} to recognize handwritten digits. The first classifier is trained on the original space $\reals^{784}$ where each feature corresponds to the intensity of one particular pixel whereas the second one is also built on $\reals^{784}$ but after transforming the space using diffusion. In Figure \ref{fig_number_classification_error} we present the error rates for SVM classification between subsets of digits which are hard to distinguish, such as 3 and 5. For these experiments, we choose 800 random samples of the digits being analyzed from the MNIST database and partition the sampled data into two halves corresponding to the training and testing data. Within the training data we perform $5$-fold cross validation to select the best combinations of the penalty parameter for the error term in the SVM objective function and the spreading parameter in the radial basis function kernel. We then train the SVM using the entire training set with the best parameter combinations and compute the accuracy of using the trained model to classify the testing data.
From the figure it is immediate that the diffusion transformation reduces the classification error, e.g. when distinguishing between 3, 5, 8, and 9, the error is reduced from $4.5\%$ to $2\%$, and when distinguishing between 1, 2, and 7 the error is reduced from $1.75\%$ to perfect attribution. Similarly, we compare the accuracy of both approaches when training a multi-class classifier to categorize among the ten possible digits. We run this experiment for a sample size of 2000 digits equally distributed across the ten possible digits of which 80\% is considered training data and the rest testing data. We follow the same training procedure described for the classification of subsets of digits. The total error obtained by the original approach is $5.75\%$ whereas by using diffusion to preprocess the data we reduce this error to $4.25\%$, i.e. a $26\%$ reduction of the error rate.

\section{Conclusion}\label{sec_conclusion}
The superposition and diffusion distances, as metrics to compare signals in networks, were introduced. Both metrics rely on the temporal heat map induced by the diffusion of signals across the network. The superposition distance quantifies the instantaneous difference between the diffused signals while the diffusion distance evaluates the accumulated effect across time. Both distances were shown to be stable with respect to perturbations in the underlying network, however, due to its closed form, the diffusion distance was found to be more suitable for implementation. We showed how both distances can be used to obtain a better classification of signals in networks both in synthetic settings as well as in a real-world classification of cancer histologies. Finally, we reinterpreted diffusion as a transformation of the feature space which can be used as a preprocessing step in learning, and illustrated its utility by classifying handwritten digits.



\begin{appendices}

\section{Proof of Theorem \ref{thm_spsnorm_stability}}\label{app_stability_sps}

The following lemma is central to the proof of Theorem \ref{thm_spsnorm_stability}.

%
\begin{lemma}\label{lemma_exponential_laplacian_doubly_stochastic}
Given the Laplacian $L$ for some undirected network, the matrix exponential of nonpositive multiples of the Laplacian $e^{-\tau L}$ with $\tau \ge 0$ is a doubly stochastic matrix.
\end{lemma}
\begin{myproof}
Since $L = D - A$, all off-diagonal components of $L$ are nonpositive, therefore $-L$ and $-\tau L$ are Metzler matrices. Since the exponentials of Metzler matrices are nonnegative \cite[Theorem 8.2]{Varga00}, we are guaranteed that all elements of $e^{-\tau L}$ are nonnegative. From the power series of matrix exponentials, we have
\begin{align}\label{eqn_proof_exp_inv_exponential_series}
      e^{-\tau L} = \sum_{k=0}^\infty \frac{1}{k!} (-\tau L)^k = 
            I - \tau L + \frac{\tau^2 L^2}{2} - \frac{\tau^3 L^3}{3!} + \cdots.
\end{align}
If we are able to show that all rows and columns of $L^k$ add up to $0$ for any integer $k \ge 1$, then we know that all rows and columns of $\sum_{k=1}^\infty (-\tau L)^k/k!$ also add up to $0$. Therefore, when we add the identity matrix to this summation to obtain the exponential $e^{-\tau L}$ as in \eqref{eqn_proof_exp_inv_exponential_series} we are guaranteed that the rows and columns sum up to $1$. Combining this with the non negativity of $e^{-\tau L}$ implies doubly stochasticity, as wanted. We now prove that all rows and columns of $L^k$ indeed add up to $0$ for any integer $k \geq 1$. First notice that for $k=1$ this is immediate since the rows and columns of the Laplacian sum up to 0 by definition. Now, consider an arbitrary matrix $B = C \, L$ obtained by left multiplying $L$ by another matrix $C$. Then, the sum of any row of $B$ is given by
\begin{equation}\label{eqn_proof_exp_inv_L_kplus1_entry}
      \sum_j B_{ij} = \sum_j \sum_m C_{im} L_{mj} =  \sum_m C_{im}  \sum_j L_{mj} = 0,
\end{equation}
where the last equality follows from the fact that $\sum_j L_{mj} = 0$ for any $m$, i.e. all rows of the Laplacian sum up to 0. Similarly, we can show that the columns of a matrix $B = L \, C$ obtained by right multiplying the Laplacian by another matrix $C$ sum up to 0. Finally, for any power $k$, the matrix $L^k = L^{k-1} \, L = L \, L^{k-1}$ can be obtained by both right or left multiplying $L^{k-1}$ by the Laplacian $L$, thus all rows and columns of $L^k$ sum up to 0 for all $k \geq 1$.
\end{myproof}

We now use Lemma \ref{lemma_exponential_laplacian_doubly_stochastic} to show Theorem \ref{thm_spsnorm_stability}.

\begin{myproof}[of Theorem \ref{thm_spsnorm_stability}]
Given the definition of $L'$, from \eqref{eqn_definition_superposition_distance} we have that
\begin{align}\label{eqn_proof_stability1_dsuper_Lprime}
      d_{\sps}^{L'}(s,r) = \int_0^\infty e^{-t} \left\| e^{-(L+E)t} (s - r) \right\|_p dt,
\end{align}
where without loss of generality we assume $\alpha = 1$. If $\alpha \neq 1$, then $\alpha L'$ defines a Laplacian and we can think of the distance $d^{\alpha L'}_{\sps}(s,r)$ where the new $\alpha$ parameter is equal to 1.
If we focus on the input norm $\| \cdot \|_p$ inside the integral in \eqref{eqn_proof_stability1_dsuper_Lprime}, we may add and subtract $e^{-Lt}(s-r)$ to obtain
\begin{align}\label{eqn_proof_stability1_norm_triangle}
      \nonumber & \Big\| e^{-(L + E)t (s - r)} \Big\|_p \!\!\!\! = \!\! \left\| \! \left(\! e^{-(L + E)t} \!\! - e^{-Lt} \right) \!\! (s \!- \!r) \! + \! e^{-Lt} (s \! - \! r)\right\|_p \\
      &\le \left\| \left(e^{-(L + E)t} - e^{-Lt} \right)(s - r)\right\|_p  
            + \left\|e^{-Lt} (s - r)\right\|_p,
\end{align}
where we used the subadditivity property of the input norm. To further bound the first term on the right hand side of \eqref{eqn_proof_stability1_norm_triangle} we apply the compatibility property of $p$-norms \eqref{eqn_definition_compatibility} followed by the subadditivity property to obtain that 
\begin{align}\label{eqn_proof_stability1_norm_split}
      \nonumber &\left\| \! \left(e^{-(L + E)t} \! - \! e^{-Lt} \right) \! (s \! - \! r)\right\|_p \!\! \le \! \left\| e^{-(L + E)t} \! - \! e^{-Lt}  \right\|_p \!\! 
            \left\| \! (s \! - \! r) \! \right\|_p \\
       &\le \left\| e^{-(L + E)t} - e^{-Lt} \right\|_p
            \left( \|s\|_p + \|r\|_p \right).
\end{align}
In order to bound the first term on the right hand side of \eqref{eqn_proof_stability1_norm_split}, we use a well-known result in matrix exponential analysis \cite{Bellman70, VanLoan77} that allows us to write the difference of matrix exponentials in terms of an integral, 
\begin{align}\label{eqn_proof_stability1_norm_integral}
      \nonumber  \Big\| e^{-(L + E)t} & - e^{-Lt}  \Big\|_p = \left\| \int_0^t e^{-L(t-\tau)} E e^{-(L+E)\tau} d\tau \right\|_p \\
      \nonumber &\le \int_0^t \left\| e^{-L(t-\tau)} E e^{-(L+E)\tau} \right\|_p d\tau \\
      & \le \! \|E\|_p \! \int_0^t \left\| e^{-L(t-\tau)}\right\|_p \left\| 
            e^{-(L+E)\tau} \right\|_p d\tau,
\end{align}
where the first inequality follows from subadditivity of the input $p$-norm and the second one from submultiplicativity \eqref{eqn_definition_submultiplicativity}. 

We now bound each of the three terms on the right hand side of \eqref{eqn_proof_stability1_norm_integral}. For the first term, $\|E\|_p \le \epsilon \|L\|_p$ by assumption. From Lemma \ref{lemma_exponential_laplacian_doubly_stochastic}, the doubly stochasticity of $e^{-L(t-\tau)}$ implies that $\| e^{-L(t-\tau)}\|_1 = \| e^{-L(t-\tau)}\|_\infty = 1$. For $p = 2$, $-L$ being negative semi-definite with largest eigenvalue at $0$ implies that the largest eigenvalue of $e^{-L(t - \tau)}$ is equal to $1$ and hence $\|e^{-L(t - \tau)}\|_2 = 1$. For the term $\left\| e^{-(L+E)\tau} \right\|_p$, notice that $L+E = L'$ is in itself a Laplacian, meaning that we can follow the aforementioned argument and upper bound this term by 1.
Substituting these bounds in \eqref{eqn_proof_stability1_norm_integral} and solving the integral yields
\begin{align}\label{eqn_proof_stability1_norm_integral_2}
     \Big\| e^{-(L + E)t} & - e^{-Lt}  \Big\|_p \leq \epsilon \|L\|_p \, t.     
\end{align}
Further substitution in \eqref{eqn_proof_stability1_norm_split} combined with the fact that $\|s\|_p \le \gamma$ and $\|r\|_p \le \gamma$, results in
\begin{align}\label{eqn_proof_stability1_integrate}
\left\|  \left(e^{-(L + E)t} -  e^{-Lt} \right)  (s  -  r) \right\|_p \le 2 \gamma \epsilon \|L\|_p \, t.
\end{align}
By substituting this result in \eqref{eqn_proof_stability1_norm_triangle} and inputing the resultant inequality in the integral in \eqref{eqn_proof_stability1_dsuper_Lprime} we conclude that
\begin{align}\label{eqn_proof_stability1_integrate}
       d_{\sps}^{L'}(s,r) \leq \int_0^\infty \! t e^{-t} 2 \gamma \epsilon \|L\|_p dt +  \int_0^\infty \!\!\! e^{-t} \left\| e^{-Lt} (s - r) \right\|_p dt.
\end{align}
Notice that the rightmost summand in \eqref{eqn_proof_stability1_integrate} is exactly equal to $d^L_{\sps}(r, s)$ [cf. \eqref{eqn_definition_superposition_distance}]. Thus, solving the integral in the first summand we get that
\begin{align}\label{eqn_proof_stability1_integrate_in_definition}
      d_{\sps}^{L'}(s,r) - d_{\sps}^L(s,r) \le 2\gamma \epsilon \|L\|_p.
\end{align}
Following the same methodology but starting from the definition of $d^L_{\sps}(s, r)$, it can be shown that
\begin{align}\label{eqn_proof_stability1_integrate_in_definition_2}
   d_{\sps}^{L}(s,r) - d_{\sps}^{L'}(s,r)  \le 2 \gamma \epsilon \|L\|_p.
\end{align}
Finally, by combining \eqref{eqn_proof_stability1_integrate_in_definition} and \eqref{eqn_proof_stability1_integrate_in_definition_2}, we obtain \eqref{eqn_thm_spsnorm_stability}, concluding the proof.
\end{myproof}

\section{Proof of Theorem \ref{thm_diffnorm_stability}}\label{app_stability_diff}

In the proof of Theorem \ref{thm_diffnorm_stability} we use two lemmas. The first one is similar to Lemma \ref{lemma_exponential_laplacian_doubly_stochastic} and shows that $(I+L)^{-1}$ is doubly stochastic.
%
\begin{lemma}\label{lemma_inverse_laplacian_doubly_stochastic}
Given the Laplacian $L$ for some undirected network, the inverse of the Laplacian plus identity matrix $(I+L)^{-1}$ is a doubly stochastic matrix.\end{lemma}
\begin{myproof}
Since all the off-diagonal entries of $I+L$ are less than or equal to zero, $I+L$ is a $Z$-matrix \cite{Young71}. Moreover, due to the fact that all eigenvalues of $I+L$ have positive real parts, $I+L$ is an $M$-matrix. Since the inverse of an $M$-matrix is elementwise nonnegative \cite{Fujimoto04}, $(I+L)^{-1}$ is a nonnegative matrix. Thus, to show doubly stochasticity, we only need to prove that all rows and columns of $(I+L)^{-1}$ add up to $1$. Denote entries in $(I+L)$ as $l_{ij}$ and in $(I+L)^{-1}$ as $a_{ij}$, from $(I+L)^{-1}(I+L) = I$, we know that for any $i$, 
\begin{align}\label{eqn_proof_exp_inv_LplusI_left}
      \sum_k a_{ik} l_{ki} &= I_{ii} = 1, \\
      \sum_k a_{ik} l_{kj} &= I_{ij} = 0,~\forall~j \ne i. \label{eqn_proof_exp_inv_LplusI_left_2}
\end{align}
Summing \eqref{eqn_proof_exp_inv_LplusI_left_2} over all $j$ yields
\begin{align}\label{eqn_proof_exp_inv_LplusI_left_sum}
      \sum_j \left( \sum_k a_{ik} l_{kj} \right) = 
            \sum_k a_{ik} \left( \sum_j l_{kj} \right) = 1. 
\end{align}
Since $\sum_j l_{kj} = 1$ for any $k$ from the definition of the matrix $(I + L)$, we know that $\sum_k a_{ik} = 1$ implying that the summation of any rows of $(I+L)^{-1}$ is $1$. Similarly, $(I+L)(I+L)^{-1} = I$ induces that the summation of all columns of $(I+L)^{-1}$ is $1$, concluding the proof.
\end{myproof}

The second lemma is a statement about the stability of inverse matrices.
%
\begin{lemma}\label{lemma_inverse_stability}
If $A$ is nonsingular and $\|A^{-1}E\|_p < 1$, then $A + E$ is nonsingular and it is guaranteed that
\begin{align}\label{eqn_lemma_inverse_stability}
      \left\| (A + E)^{-1} - A^{-1} \right\|_p \le
            \frac{\|E\|_p \| A^{-1} \|_p^2}{1 - \|A^{-1}E\|_p}.
\end{align}
\end{lemma}
\begin{myproofnoname}
See \cite[Theorem 2.3.4]{Golub89}.
\end{myproofnoname}

We now use Lemmas \ref{lemma_inverse_laplacian_doubly_stochastic} and \ref{lemma_inverse_stability} to show Theorem \ref{thm_diffnorm_stability}.

\begin{myproof}[of Theorem \ref{thm_diffnorm_stability}]
Given the definition of $L'$, from \eqref{eqn_definition_diffusion_distance_2} we have that
\begin{align}\label{eqn_proof_thm_diffnorm_stability_dfn}
      d_{\diff}^{L'}(s,r) = \left\| (I + L + E)^{-1} (s-r) \right\|_p.
\end{align}
As in the proof of Theorem \ref{thm_spsnorm_stability}, we can assume that $\alpha=1$ without loss of generality.
Subtracting and adding $(I + L)^{-1} (s-r)$ from \eqref{eqn_proof_thm_diffnorm_stability_dfn} and applying the subadditivity property of the $p$-norm implies
\begin{align}\label{eqn_proof_thm_diffnorm_stability_addsubstract}
      \nonumber d_{\diff}^{L'}(s,r) & \le 
            \left\| \left( (I + L + E)^{-1} - (I + L)^{-1}\right)(s- r)\right\|_p\\
      &~~~~~~~~~~ + \left\| (I + L)^{-1} (s-r)\right\|_p,
\end{align}
where the second term in the sum is exactly $d_{\diff}^L(s,r)$ [cf. \eqref{eqn_definition_diffusion_distance_2}]. Therefore we may write
\begin{align}\label{eqn_proof_thm_diffnorm_stability_difference}
   d_{\diff}^{L'}(s,r) \! - \!d_{\diff}^{L}(s,r)  \! \leq \! \big\|  \big( (I \! + \! L \! + \! E)^{-1} \! - \! (I \! + \! L)^{-1}\big)(s- r) \big\|_p \!.
\end{align}
By applying compatibility of $p$-norms \eqref{eqn_definition_compatibility} followed by the subadditivity property we obtain that 
\begin{align}\label{eqn_proof_thm_diffnorm_stability_split_norms}
       d_{\diff}^{L'}(s,r) - d_{\diff}^{L}(s,r) ~~~& \\
            \nonumber \le \big\| \big( (I + L + E)^{-1} &- (I + L)^{-1}\big)\big\|_p
                  \left\| (s- r)\right\|_p \\
       \le \big\| \big( (I + L + E)^{-1} &- (I + L)^{-1}\big)\big\|_p
                  \left( \|s\|_p + \|r\|_p \right) \nonumber
\end{align}
Given that $I+L$ is nonsingular we have to show that $\|(I+L)^{-1}E\|_p < 1$ in order to be able to apply Lemma \ref{lemma_inverse_stability} with $A = (I+L)$ and further bound \eqref{eqn_proof_thm_diffnorm_stability_split_norms}.

Due to doubly stochasticity [cf. Lemma \ref{lemma_inverse_laplacian_doubly_stochastic}], we have that $\|(I+L)^{-1}\|_1 = \|(I+L)^{-1}\|_\infty = 1$. Moreover, $\|(I+L)^{-1}\|_2 = 1$ comes from the fact that the smallest eigenvalue of $(I+L)$ and hence the largest eigenvalue of $(I+L)^{-1}$ is equal to $1$. Consequently, we may write
\begin{align}\label{eqn_proof_thm_diffnorm_stability_able_to_apply_lemma}
      \|(I+L)^{-1}E\|_p \le \|(I+L)^{-1}\|_p \|E\|_p <1,
\end{align}
for $p \in \{1, 2, \infty\}$, as wanted, where the first inequality follows from submultiplicativity \eqref{eqn_definition_submultiplicativity}. Hence, applying Lemma \ref{lemma_inverse_stability} with $A = (I+L)$ yields
\begin{align}\label{eqn_proof_thm_diffnorm_stability_apply_lemma}
      \big\|  (I + L + E)^{-1} - (I + L)^{-1} \big\|_p \le 
            \frac{\|E\|_p \| (I+L)^{-1} \|_p^2}{1 - \|(I+L)^{-1}E\|_p}.
\end{align}
Recalling that $\|(I+L)^{-1}\|_p = 1$ for any $p \in \{1, 2, \infty\}$ allows us to further bound \eqref{eqn_proof_thm_diffnorm_stability_apply_lemma} to obtain
\begin{align}\label{eqn_proof_thm_diffnorm_stability_apply_lemma_1}
    \big\|  (I + L + E)^{-1} \! - \! (I + L)^{-1} \big\|_p \le \frac{\|E\|_p}{1 - \|E\|_p} \! \le \! \frac{\epsilon\|L\|_p}{1 - \epsilon\|L\|_p},
\end{align}
where we used that $\|E\|_p \le \epsilon \|L\|_p < 1$ for the last inequality.

Utilizing the Taylor series of $1 / (1 - \epsilon\|L\|_p)$ and substituting \eqref{eqn_proof_thm_diffnorm_stability_apply_lemma_1} into \eqref{eqn_proof_thm_diffnorm_stability_split_norms} combined with the fact that $\|s\|_p \le \gamma$ and $\|r\|_p \le \gamma$ we have that
\begin{align}\label{eqn_proof_thm_diffnorm_stability_final}
   d_{\diff}^{L'}(s,r) - d_{\diff}^{L}(s,r)  \le
            \sum_{n = 1}^\infty 2\gamma ( \epsilon \|L\|_p)^n 
       = 2 \gamma \|L\|_p \epsilon + o(\epsilon).
\end{align}
In a similar manner but starting from the definition of $d^L_{\diff}(s, r)$, it can be shown that
\begin{align}\label{eqn_proof_thm_diffnorm_stability_final_2}
   d_{\diff}^{L}(s,r) - d_{\diff}^{L'}(s,r)  \le 2 \gamma \|L\|_p \epsilon + o(\epsilon).
\end{align}
Finally, by combining \eqref{eqn_proof_thm_diffnorm_stability_final} and \eqref{eqn_proof_thm_diffnorm_stability_final_2}, we obtain \eqref{eqn_thm_diffnorm_stability} and the proof concludes.
\end{myproof}
\end{appendices}

\urlstyle{same}
\bibliographystyle{IEEEtran}
\bibliography{diffusion_biblio.bib}

\end{document}